\documentclass{article}

\usepackage{microtype}
\usepackage{graphicx}
\usepackage[dvipsnames]{xcolor}
\usepackage{booktabs} 
\usepackage{float}
\usepackage{multirow}
\usepackage{multicol}
\usepackage{enumitem}
\usepackage{makecell}
\usepackage{siunitx}
\usepackage{graphicx}
\usepackage{subcaption}
\usepackage[normalem]{ulem}
\usepackage{caption}
\usepackage{stfloats}
\usepackage{longtable}
\usepackage{array}
\usepackage{makecell}

\newcolumntype{M}{>{$}c<{$}}
\usepackage{hyperref}

\usepackage[accepted]{icml2025}

\usepackage{amsmath}
\usepackage{amssymb}
\usepackage{mathtools}
\usepackage{amsthm}
\usepackage{pifont}
\newcommand{\cmark}{\ding{51}}%
\newcommand{\xmark}{\ding{55}}%

\newcommand{\name}{MindLLM}
\usepackage{subcaption}
\usepackage[capitalize,noabbrev]{cleveref}
\usepackage{tabularx}
\usepackage{adjustbox}
\usepackage{makecell}
\usepackage[most]{tcolorbox}

\theoremstyle{plain}

\theoremstyle{definition}

\theoremstyle{remark}

\usepackage[textsize=tiny]{todonotes}
\newcommand{\papertitle}{\name{}: A Subject-Agnostic and Versatile Model for fMRI-to-Text Decoding}

\icmltitlerunning{\papertitle{}}

\begin{document}

\twocolumn[
\icmltitle{\raisebox{-13pt}{\includegraphics[width=0.06\textwidth]{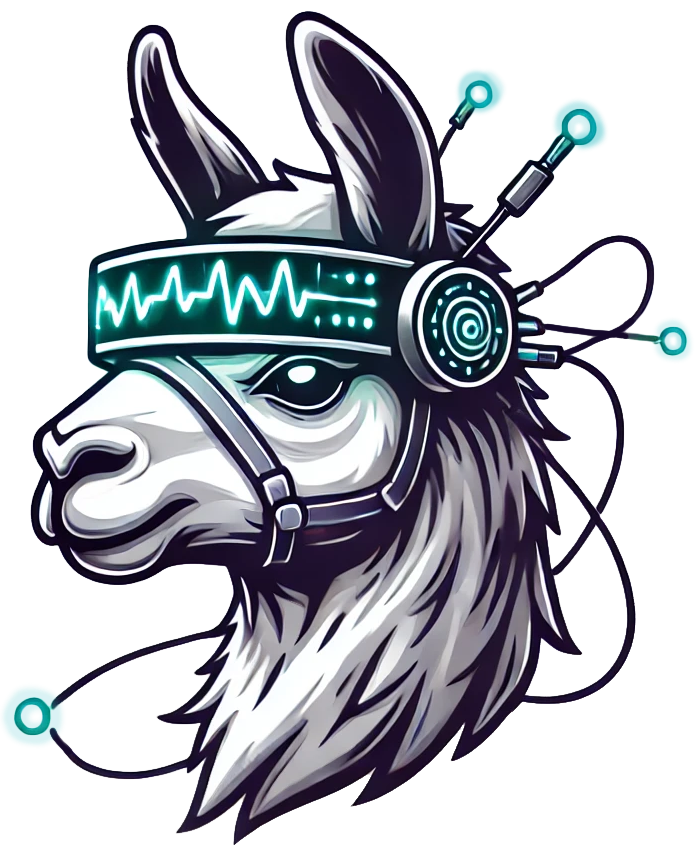}}\hspace{5pt}\name{}:\\[4pt]A Subject-Agnostic and Versatile Model for fMRI-to-Text Decoding}



\icmlsetsymbol{equal}{*}

\begin{icmlauthorlist}
\icmlauthor{Weikang Qiu}{yale}
\icmlauthor{Zheng Huang}{equal,dartmouth}
\icmlauthor{Haoyu Hu}{equal,cambridge}
\icmlauthor{Aosong Feng}{yale}
\icmlauthor{Yujun Yan}{dartmouth}
\icmlauthor{Rex Ying}{yale}
\end{icmlauthorlist}

\icmlaffiliation{yale}{Yale University}
\icmlaffiliation{dartmouth}{Dartmouth College}
\icmlaffiliation{cambridge}{University of Cambridge}

\icmlcorrespondingauthor{Weikang Qiu}{weikang.qiu@yale.edu}

\icmlkeywords{Machine Learning, ICML}

\vskip 0.3in
]



\printAffiliationsAndNotice{\icmlEqualContribution} 

\begin{abstract}
Decoding functional magnetic resonance imaging (fMRI) signals into text has been a key challenge in the neuroscience community, with the potential to advance brain-computer interfaces and uncover deeper insights into brain mechanisms. However, existing approaches often struggle with suboptimal predictive performance, limited task variety, and poor generalization across subjects. In response to this, we propose \name{}, a model designed for subject-agnostic and versatile fMRI-to-text decoding. \name{} consists of an fMRI encoder and an off-the-shelf LLM. The fMRI encoder employs a neuroscience-informed attention mechanism, which is capable of accommodating subjects with varying input shapes and thus achieves high-performance subject-agnostic decoding. Moreover, we introduce Brain Instruction Tuning (BIT), a novel approach that enhances the model's ability to capture diverse semantic representations from fMRI signals, facilitating more versatile decoding. We evaluate \name{} on comprehensive fMRI-to-text benchmarks. Results demonstrate that our model outperforms the baselines, improving downstream tasks by $12.0\%$, unseen subject generalization by $24.5\%$, and novel task adaptation by $25.0\%$. Furthermore, the attention patterns in \name{} provide interpretable insights into its decision-making process. Code is available at \url{https://github.com/Graph-and-Geometric-Learning/MindLLM}.


\end{abstract}

\section{Introduction}
\begin{figure}
    \centering
    \includegraphics[width=\linewidth]{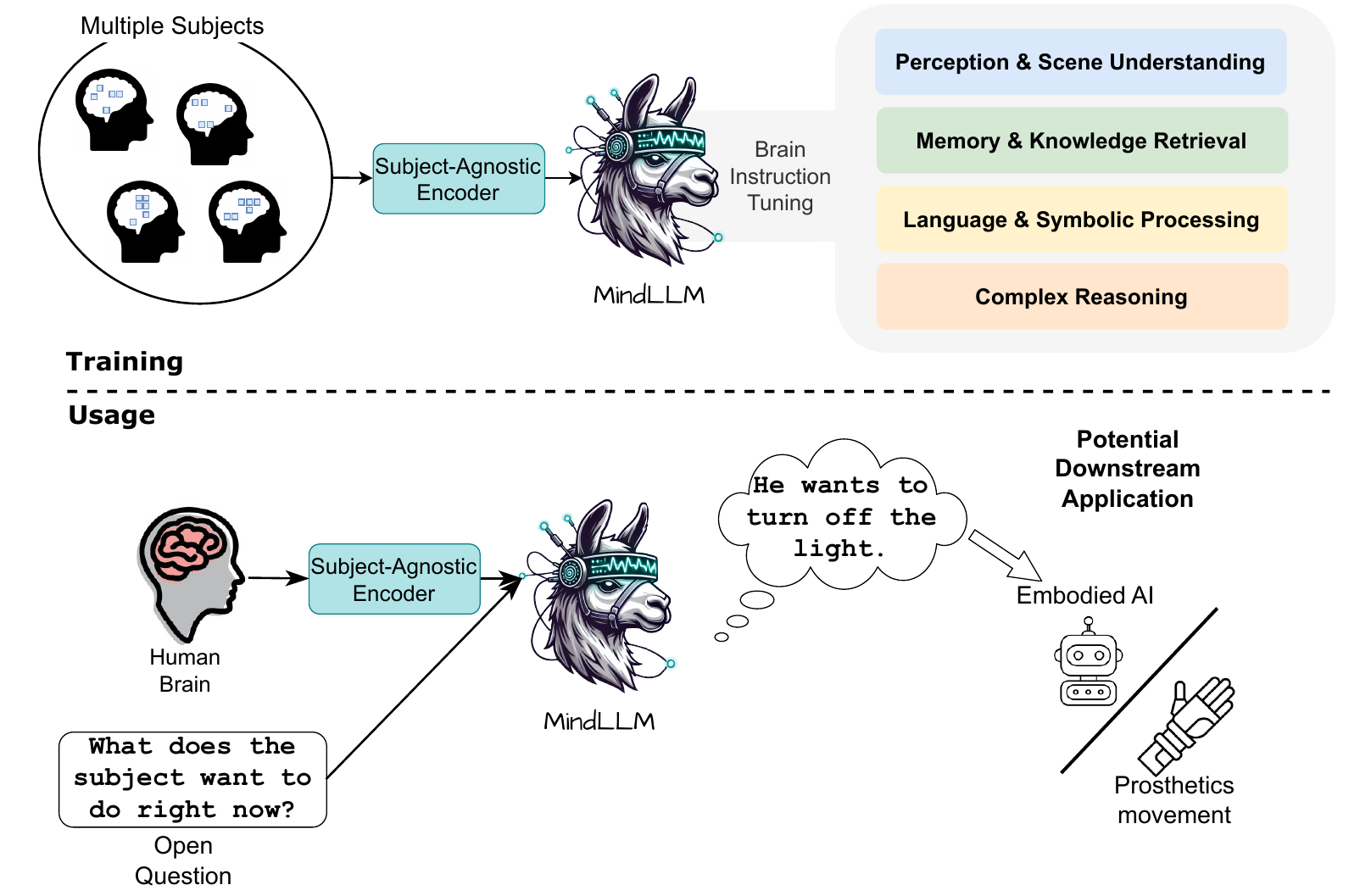}
    \vspace{-2em}
    \caption{The overview of our method. \name{} is equipped with a subject-agnostic fMRI encoder and an off-the-shelf LLM. \name{} is trained on multiple subjects with varying input shapes and an instruction-tuning dataset, aiming to encode different facets of semantic information in fMRI. After training, \name{} is capable of various text decoding tasks. One application is that the decoded contents can be used to achieve neural control of existing systems that are not designed for it.}
    \label{fig:teaser}
    \vspace{-1.5em}
\end{figure}

Decoding human brain activity (fMRI) to text has sparked significant interest within the neuroscience community \cite{xia2024umbrae,chen2023mindgpt,luo2023brainscuba,hmamouche2024multimodal}. The ability to translate brain activity patterns into natural language carries both academic and societal importance. For neuroscientists, it provides deeper and novel insights into cognition, behavior, and consciousness \cite{qiu2023learning,luo2023brainscuba}. On a societal level, it presents opportunities for medical applications and improves human-computer interaction (HCI) \cite{bernal2022brain,du2022fmri}. For example, for individuals with speech impairments, this technology could restore communication capabilities, enabling them to express their thoughts effortlessly \cite{card2024accurate}. Moreover, as shown in Figure~\ref{fig:teaser}, it benefits healthy individuals by allowing neural control of digital devices, such as embodied AIs or prosthetic limbs, allowing for more intuitive and precise movements.

Despite its potential, decoding brain activity to language still faces significant challenges. One major obstacle is the need for versatile decoding tailored to specific applications. For example, decoding may aim to translate a subject's movement intention to control a prosthesis, or to interpret abstract thoughts or memories. Traditional models fail to accommodate such diverse requirements. To address this, UMBRAE \cite{xia2024umbrae} integrates a Visual Language Model (VLM) \cite{chen2023shikra} 
 and learns to map from fMRI data to corresponding stimulus images. While this approach achieves versatility to some extent, it remains constrained to tasks directly tied to the current stimulus image and cannot address broader tasks, such as retrieving memories of past visual experiences.

Another critical challenge lies in designing a unified and subject-agnostic architecture. Current methods of brain multimodal decoding mostly rely on a preprocessing step: selecting responsive voxels by comparing the task-based fMRI to the resting-state fMRI. The selection typically results in higher performance compared to using whole-brain data. However, the varying number and irregular spatial distribution of selected voxels across subjects pose significant challenges for developing a unified architecture. To this end, recent studies \cite{wang2024mindbridge,wang2024unibrain} have proposed pooling or sampling voxels to standardize input dimensions. However, as illustrated in Figure~\ref{fig:arch}, these methods still suffer from the loss of spatial information and uneven representations of certain areas, ultimately compromising performance.

\noindent\textbf{Present Work} Here we present \name{}, a subject-agnostic and versatile model for fMRI-to-text decoding. Our approach consists of a subject-agnostic fMRI encoder and an off-the-shelf LLM. The subject-agnostic fMRI encoder incorporates a neuroscience-informed attention layer with learnable \textit{queries}, enabling dynamic feature extraction by leveraging both spatial information and neuroscientific priors of voxels, thereby significantly enhancing prediction accuracy. The design of \textit{values} and \textit{keys} separates the voxel's functional information-which is largely consistent across individuals-from its fMRI value, allowing the model to benefit from shared priors across subjects and enhancing generalization to novel subjects. Moreover, to address the challenge of versatile decoding, we propose Brain Instruction Tuning (BIT). BIT trains the model using a diverse dataset that employs images as intermediaries, encompassing tasks designed to capture diverse aspects of semantic information encoded in fMRI data, including perception \& scene understanding, memory \& knowledge retrieval, language \& symbolic processing, and complex reasoning. Figure~\ref{fig:teaser} illustrates the corresponding components.

We evaluate our model on comprehensive benchmarks. Results reveal it outperforms baselines with $12.0\%$ average improvement in various downstream tasks and $24.5\%$ improvement in generalization on unseen subjects. Additionally, we show that our model adapts effectively to novel tasks, demonstrating high customizability and flexibility in real-world applications. Furthermore, our analysis of attention weights offers valuable insights into the working mechanism of our fMRI encoder.

\section{Related Works}
\noindent\textbf{Brain-Conditioned Text Generation}

\begin{figure}
    \centering
    \includegraphics[width=\linewidth]{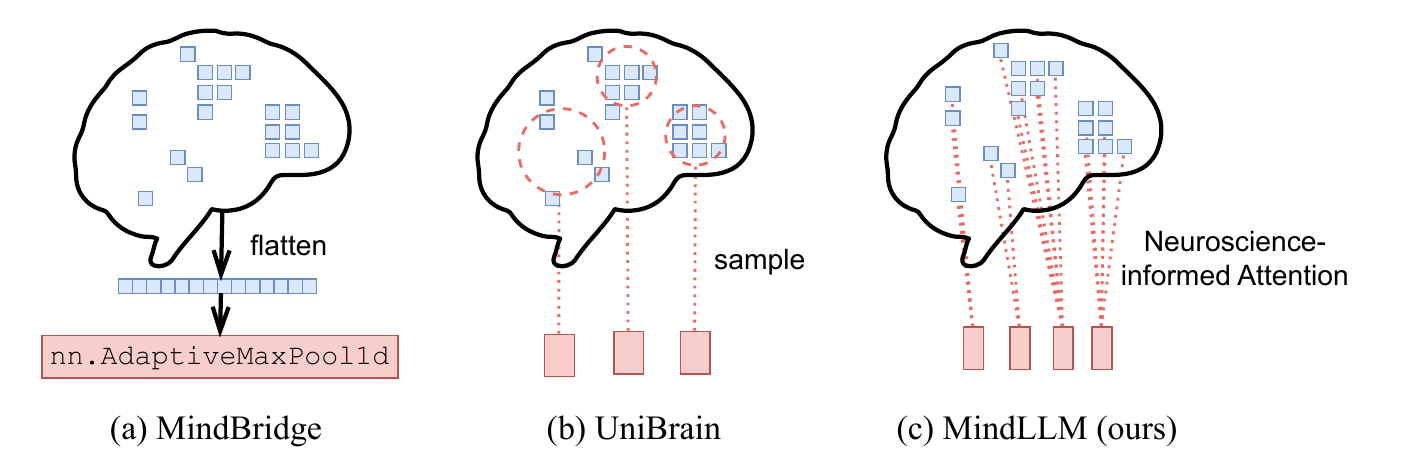}
    \vspace{-2em}
    \caption{Comparison between our model and previous unified models. MindBridge \cite{wang2024mindbridge} flattens the voxels and adaptively pools them to a fixed dimension, which overlooks the rich information in positions. UniBrain~ \cite{wang2024unibrain} uniformly samples a subset of voxels and aggregates their neighbors. Different from these methods, we propose neuroscience-informed attention, where each query token attends to all voxels, which minimizes potential information loss in pooling or sampling.}
    \label{fig:uni_compare}
    \vspace{-1em}
\end{figure}
This line of research mostly focuses on decoding perceived visual stimuli into natural language from fMRI signals. MindGPT~\cite{chen2023mindgpt}, UniBrain\cite{mai2023unibrain} and BrainCap \cite{ferrante2023brain} employ an fMRI encoder guided by CLIP~\cite{radford2021learning} and use a language model \cite{radford2019language,wang2022git} to decode natural language from the encoded representations. BrainChat~\cite{huang2024brainchat} utilizes multiple pretraining strategies~\cite{devlin2018bert, he2022masked,yu2022coca} to align fMRI with image and text embeddings. These methods fall short in performance and versatility. UMBRAE~\cite{xia2024umbrae} proposes to learn a mapping from fMRI to stimulus images, which later serves as a proxy input for an off-the-shelf visual language model (VLM). Although they achieve performance improvements, the strategy prevents the model from performing tasks that are not directly related to the stimulus images (e.g., answering memory-related questions). In contrast, our end-to-end Brain Instruction Tuning (BIT) ensures seamless and versatile fMRI-to-text decoding, offering the potential to tackle tasks beyond vision-related ones.

\noindent\textbf{Cross-subjects Decoding}
In voxel-level machine learning for brain decoding, the number of voxels varies between subjects \cite{allen2022massive}. Most prior works \cite{scotti2024reconstructing,scotti2024mindeye2} use an MLP for each subject individually. However, due to the fixed input size required by MLP architectures, these models cannot handle varying input shapes. As illustrated in Figure~\ref{fig:uni_compare}, MindBridge \cite{wang2024mindbridge} proposed to use an adaptive max pooling layer to standardize the input shapes. However, unlike images, which are considered translation invariance, positions in fMRI carry specific bio-informative significance that pooling operations may overlook. UniBrain \cite{wang2024unibrain} proposed to sample groups of voxels. Such a sampling strategy, on the one hand, may lead to information loss if some voxels are not included in any group. On the other hand, the irregular spatial distribution of 3D voxels with varying density and curvature may result in underrepresentation or overrepresentation of certain areas. Different from these methods, our model employs a neuroscience-informed attention mechanism that accounts for every single voxel while preserving their bio-informative positional information, ensuring a more comprehensive and precise representation. 

\noindent\textbf{Multi-Modal Large Language Model}
Aiming to augment the perceptual capacities of Large Language Models (LLMs), there has been a growing interest in extending them to handle multiple modalities within a unified model. Numerous studies have attempted to incorporate modalities such as images \cite{alayrac2022flamingo,zhang2023internlm,wang2023cogvlm}, videos \cite{cheng2024videollama, kondratyuk2023videopoet,zhang2023video}, and point clouds \cite{xu2023pointllm,qi2025shapellm}. 
OneLLM~\cite{han2024onellm} stands out by aligning eight different modalities, including fMRI, with language. However, their approach employs an individual convolution network for each subject instead of a unified architecture for fMRI encoding across subjects, which restricts its applicability to new subjects in real-world scenarios. Furthermore, the approach solely relies on captions as textual annotations, which limits the model's capability for versatile fMRI decoding. 

\paragraph{Multimodal fMRI Decoding}
Beyond text, recent studies have explored decoding brain signals into other modalities, including images \cite{scotti2024mindeye2, wang2024mindbridge,chen2023seeingbrainconditionaldiffusion}, videos \cite{sun2024neurocinedecodingvividvideo,chen2023cinematicmindscapeshighqualityvideo}, speech \cite{chen2024open} and music \cite{denk2023brain2musicreconstructingmusichuman}, as well as in reverse directions \cite{toneva2019interpreting}. The encoder architectures in these works typically either adopt transformers for time-series data, which is distinct from our voxel-based setting, or fall into one of the architectures we have discussed in previous paragraphs. Therefore, focusing on text generation, our work improves the fMRI encoder architecture for voxel-level inputs, and designs novel objectives (i.e., brain instruction tuning) for our target modality.

\section{Method}
\label{sec:method}
In this section, we propose a neuroscience-informed fMRI encoder designed to achieve high-performance, subject-agnostic decoding. To further enable versatile decoding, we introduce the construction of a brain instruction tuning dataset, which captures diverse semantic representations encoded in fMRI data.

\subsection{Method Overview}
As illustrated in Figure~\ref{fig:arch}, our model consists of an fMRI encoder $f_\theta$ and an off-the-shelf LLM. In practice, we use Vicuna-7b \cite{zheng2023judging} as our LLM to maintain consistency with our baseline \cite{xia2024umbrae}. For each sample, let $\boldsymbol{v} = [v_1, v_2, \cdots, v_N]\in \mathbb{R}^N$ be the fMRI signals of input voxels, where $N$ is the number of voxels. Note that $N$ varies between different subjects, ranging from $12,682$ to $17,907$ in the dataset we use \cite{allen2022massive}.

The fMRI encoder $f_\theta$, featuring a neuroscience-informed attention layer, encodes $\boldsymbol{v}$ to fMRI tokens $X_v = [\boldsymbol{x}_{v,1}, \boldsymbol{x}_{v,2}, \cdots, \boldsymbol{x}_{v,L}] \in \mathbb{R}^{d\times L}$, where $L$ is the number of tokens and $d$ is the dimension of token embeddings. We then prepend these learned fMRI tokens to the language tokens in the BIT dataset we propose.

\subsection{fMRI Encoder}
As mentioned before, currently most models for fMRI decoding can not handle varying input shapes and are not subject-agnostic, with only a few exceptions \cite{wang2024unibrain}. However, these exceptions still suffer from information loss and uneven representations of certain brain areas. To this end, we propose a novel neuroscience-informed attention mechanism to accommodate varying voxel numbers across subjects, enabling a subject-agnostic encoding strategy. Below we talk about the design of \textit{queries} $\{\boldsymbol{q}_i\}$, \textit{keys} $\{\boldsymbol{k}_i\}$ and \textit{values} $\{\boldsymbol{v}_i\}$ in the attention layer. For \textit{values}, we directly use the fMRI signal of each voxel, which means $\boldsymbol{v_i} = v_i \in \mathbb{R}$. Making each voxel a \textit{value} token maximally prevents information loss compared to pooling- \cite{wang2024mindbridge} or sampling-based \cite{mai2023unibrain} methods. The \textit{queries} are randomly initialized and learnable. We expect each \textit{query} to represent a certain pattern of the brain (refer to visualizations in Section \ref{sec:vis}). The design of \textit{keys} will be discussed below.

\noindent\textbf{Exclude fMRI values from \textit{keys}}
The vanilla cross attention \cite{zhu2020deformable,vaswani2017attention} derives both \textit{keys} and \textit{values} from the same input source. However, we found this would lead to poor performance in fMRI. We argue the reason: different from images or text, which are usually considered translation-invariant, the positions of voxels carry specific brain \textit{functional information}, as voxels in different areas are associated with distinct brain functions. Consequently, a voxel's position alone can theoretically serve as effective \textit{keys} for attention weight computation \cite{mcmanus2002right,zhang2023abnormal,liu2024relationships}. Including fMRI values into \textit{keys}, however, introduces additional noise instead of valuable information, thus resulting in poorer performance. Moreover, since brain regions tend to serve similar functions across individuals, decoupling voxel positions from fMRI signals can facilitate the sharing of priors across subjects, potentially improving generalization to unseen subjects.

In light of this, instead of the vanilla cross attention, which derives the \textit{keys} and \textit{values} from the same inputs, we exclude the fMRI value of each voxel and use its positional information alone as its \textit{key} embedding. The positional information is encoded from the coordinates of each voxel, i.e. $\boldsymbol{k}_i^{\text{pos}} = \operatorname{PE}(\boldsymbol{c}_i)$ for the $i$-th voxel, where $\boldsymbol{c}_i \in \mathbb{R}^3$ denotes the coordinates of the voxel. In practice, we use the Fourier positional encoding proposed in \cite{tancik2020fourier} due to its superiority in encoding coordinate information.

\noindent\textbf{Incorporation of Brain Parcellations}
While positional encoding alone improves performance, it lacks inherent neuroscientific grounding, potentially making it challenging for the model to efficiently learn representations aligned with established principles of brain function. To overcome this, we incorporate existing brain region parcellations \cite{glasser2016multi,rolls2020automated} into the \textit{key} embeddings. Formally, given a parcellation $\mathcal{P}$, with regions indexed by $1, \cdots, N_\mathcal{P}$. Let $\mathcal{P}(i) \in [1, 2, \cdots, N_\mathcal{P}]$ be the region that the $i$-th voxel belongs to, and $E[\mathcal{P}(i)] \in \mathbb{R}^d$ be the corresponding learnable embedding of the region, which will be incorporated in the \textit{key} embeddings as $\boldsymbol{k}_i^{\text{reg}, \mathcal{P}} = E[\mathcal{P}(i)] \in \mathbb{R}^d$.

\begin{figure}
    \centering
    \includegraphics[width=\linewidth]{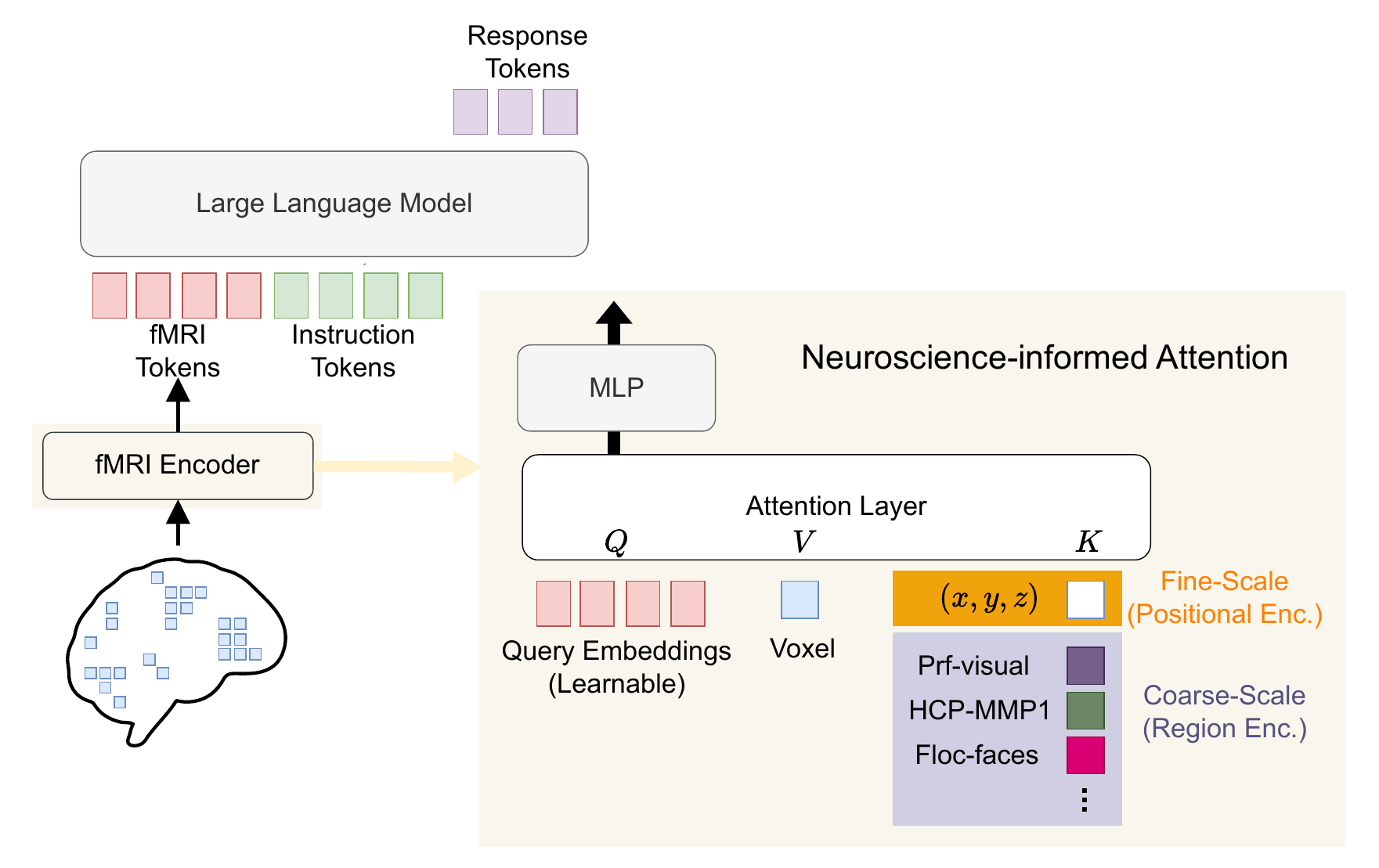}
    \caption{Model Architecture. The fMRI encoder maps fMRI to a series of fMRI tokens through our proposed neuroscience-informed attention. The large language model, with both fMRI and text tokens, will be trained by brain instruction tuning.}
    \label{fig:arch}
    \vspace{-1em}
\end{figure}

\noindent\textbf{Combining Multiple Parcellations}
It is crucial to choose an appropriate brain region parcellation. Previous region-based methods \cite{qiu2023learning,li2021braingnn, kan2022brain} can usually only choose one arbitrarily. In contrast, our model design allows us to combine multiple parcellations $\mathcal{P}^1, \mathcal{P}^2, \cdots$ by concatenating their respective region encodings to the \textit{key} embeddings. In conclusion, the final \textit{key} embeddings are the concatenation of the positional encoding and multiple region encodings,
\begin{equation}
    \boldsymbol{k}_i = \boldsymbol{k}_i^\text{pos} \| \boldsymbol{k}_i^{\text{reg}, \mathcal{P}^1} \|  \boldsymbol{k}_i^{\text{reg}, \mathcal{P}^2} \| \cdots
\end{equation}
where $\|$ denotes the concatenation operation. This process is illustrated in Figure~\ref{fig:arch}'s lower right part.

The positional and region encodings complement each other: The region encodings serve as coarse-scale features, providing a neuroscientific-grounded basis, while the fine-scale positional encoding allows our model to learn finer-grained information directly from the data.

This attention design separates a voxel's \textit{functional information}—which is largely consistent across individuals—from its fMRI value, thereby enhancing generalization. Instead of relying on pooling or sampling, the attention mechanism employs learnable aggregation, while the integration of positional encoding and neuroscientifically informed region encodings further ensures high performance.

After the attention layer, we obtain the hidden representations $\boldsymbol{z}_q \in \mathbb{R}^{N_q} $ where $N_q$ is the number of query embeddings. We then employ an MLP and a reshape operation to map the hidden representations to $L$ fMRI tokens, i.e., $   X_v = \operatorname{reshape}\left( \operatorname{MLP}(
    \{\boldsymbol{z}_q\}
    ) \right) \in \mathbb{R}^{L \times d}$.

The process of the fMRI encoder is illustrated in Figure~\ref{fig:arch}. The obtained fMRI tokens are then prepended to the language tokens in conversations.

In addition, we provide a side-by-side comparison between our encoder and existing ones in Appendix~\ref{app:arch_compare}.

\subsection{Brain Instruction Tuning (BIT)}
To enable versatile fMRI-to-text decoding, an appropriate BIT dataset is required, yet no such dataset currently exists. To bridge this gap, we construct one based on the fact: MSCOCO images \cite{chen2015microsoft} serve as stimuli for fMRI recordings in the fMRI study \cite{allen2022massive}, and an abundance of datasets provide text annotations (e.g., VQA) for MSCOCO images. Using the images as intermediaries, we select those relevant to brain functions and pair the fMRI data with corresponding text annotations. For example, given an image of a billboard with annotated textual content, we can reasonably infer that when a subject perceives textual information (e.g., contents on the billboard), corresponding representations are encoded in the brain. This suggests the possibility of extracting such information from fMRI signals. We select datasets to fulfill various purposes, enabling the model to capture diverse aspects of semantic information embedded in fMRI signals, including visual perception \& scene understanding, language \& symbolic processing, memory \& knowledge retrieval and complex reasoning, which are considered among most fundamental and essential properties of human brains \cite{robertson2002memory,stenning2012human,wade2013visual,friederici2017language}.

\begin{figure}[h]
    \centering
    \includegraphics[width=\linewidth]{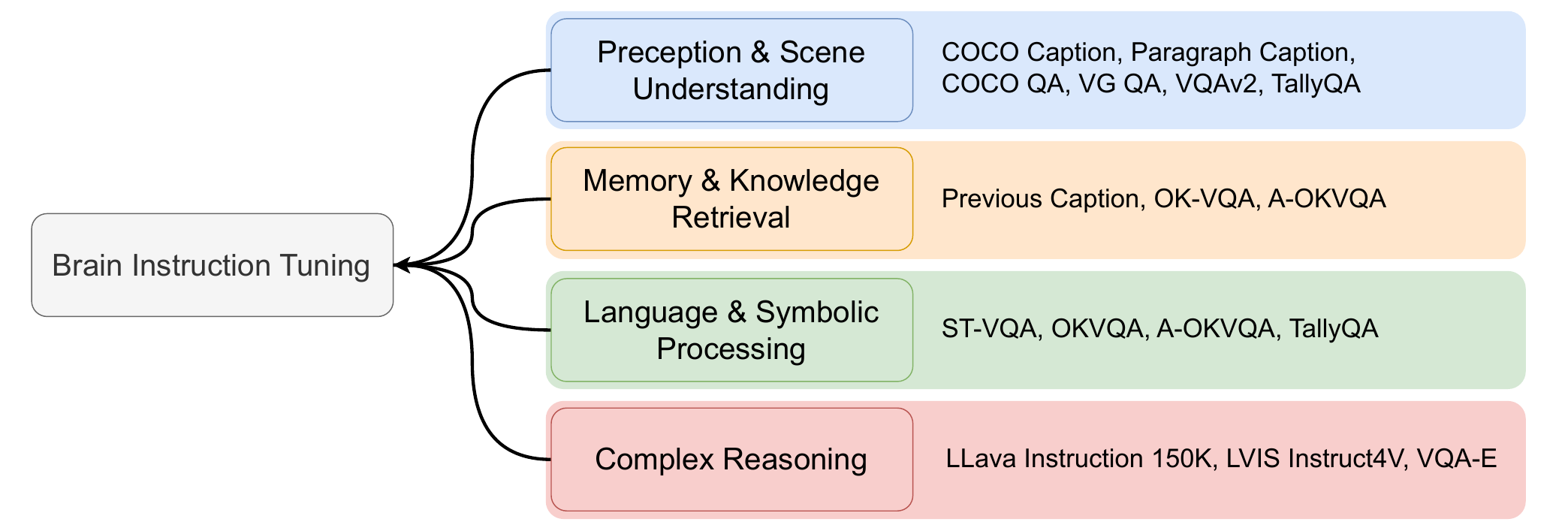}
\vspace{-1.8em}
    \caption{Dataset Taxonomy in Brain Instruction Tuning.}
    \label{fig:bit}
\end{figure}

\noindent\textbf{Perception \& Scene Understanding} As illustrated in Figure~\ref{fig:bit}, we begin by using caption tasks at both coarse and fine-grained levels to train the model’s ability to understand and summarize what the subject perceives visually \cite{chen2015microsoft,krause2017hierarchical}. Additionally, we incorporate QA tasks \cite{ren2015exploring,krishna2017visual,acharya2019tallyqa} to enhance the model's ability to retrieve and reason about visually perceived content.

\noindent\textbf{Memory \& Knowledge Retrieval} To go beyond tasks directly related to present visual perception, we construct the \emph{previous captioning} task, a memory-oriented task that challenges the model to caption images that the subject previously viewed, simulating memory recall processes. Furthermore, we aim to encode knowledge structures in human brains. The OK-VQA \cite{marino2019ok} and A-OKVQA \cite{schwenk2022okvqa} datasets include questions requiring external knowledge that is not present in the image but resides in human brains. For example, A photo of a hydrant may prompt the answer "firetruck," even though the firetruck is absent in the image. This association also reflects the way human cognition operates through a network of interconnected meanings, where one concept unconsciously triggers another. Such a process, which is called "slippage of the signifier" \cite{lacan2001ecrits, lacan1988seminar, miller2018four}, highlights the symbolic processes through which the brain constructs and retrieves meaning. 

\noindent\textbf{Language \& Symbolic Processing} In addition to the aforementioned OK-VQA and A-OKVQA datasets, which are also related to symbolic process, we further combine datasets of text recognition \cite{biten2019scene} and numerical reasoning \cite{acharya2019tallyqa} to facilitate this aspect.

\noindent\textbf{Complex Reasoning} Finally, we try to approximate the reasoning process that happens in human brains with datasets \cite{liu2023visual,wang2023see,li2018vqa} that require intricate logical and inferential processes. We expect these datasets to challenge the model to extract the reasoning process, drawing upon both visual understanding and abstract problem-solving, thus bridging perception, memory, and knowledge into a cohesive cognitive framework.

We ended up with a brain instruction tuning dataset consisting of $980,610$ conversations associated with fMRI recordings from $15$ datasets. Appendix~\ref{app:dataset} lists the instructions and other details for each dataset. The instruction tuning enables versatile fMRI-to-text decoding. In particular, the introduction of tasks like \textit{previous caption} empowers the model to perform a broader range of tasks beyond vision-related ones, which the previous model \cite{xia2024umbrae} fails.

\begingroup
\sisetup{
  table-format=2.2,  
  table-align-text-pre=false,
  propagate-math-font=true,
  table-number-alignment=center,
  detect-weight=true,detect-inline-weight=math
}
\def\Uline#1{#1\llap{\uline{\phantom{#1}}}}
\begin{table*}[bp]
    \centering
    \caption{Results of brain captioning. The CIDEr metric is scaled by a factor of 100 for consistency with Table~\ref{tab:caption} and baselines.}
    \label{tab:caption}
    \vskip 0.1in
    \resizebox{\linewidth}{!}{
    \begin{tabular}{lcSSSSSSSS}
    \toprule
   {Method} & {\makecell{subject\\agnostic}}  &{{BLEU-1} $\uparrow$} & {BLEU-2 $\uparrow$} & {BLEU-3 $\uparrow$} & {{BLEU-4} $\uparrow$} &{METEOR $\uparrow$}&{ROUGE $\uparrow$}& {CIDEr $\uparrow$}&{SPICE $\uparrow$}\\
    \midrule
    SDRecon \cite{takagi2023high}    & {\xmark} &36.21 & 17.11 & 7.22 & 3.43   &10.03&  25.13&13.83 &5.02 \\
    OneLLM  \cite{han2024onellm}  & {\xmark} &47.04 & 26.97 & 15.49 & 9.51   &13.55&  35.05&22.99 & 6.26\\
    UniBrain \cite{mai2023unibrain}   & {\xmark} & {$-$}   & {$-$}    & {$-$}  & {$-$}     &16.90&  22.20& {$-$} & {$-$}\\
    BrainCap \cite{ferrante2023brain}  & {\xmark} &55.96 & 36.21 & 22.70 & 14.51   &16.68& 40.69&41.30 & 9.06\\
     BrainChat \cite{huang2024brainchat} & {\xmark}   &52.30& 29.20& 17.10& 10.70 &14.30& 45.70&26.10 & {$-$}\\
    UMBRAE \cite{xia2024umbrae}    & {\xmark} & 59.44& 40.48& 27.66&19.03&19.45&  43.71&\bfseries 61.06&\bfseries 12.79\\
UniBrain \cite{wang2024unibrain}  & {\cmark} & 59.08 & 39.64 & 26.36 & 17.68 & 17.49 & 43.48 & 48.20 & 9.38\\
    \name{} (Ours)  & {\cmark} & \bfseries 61.75 &  \bfseries42.84 & \bfseries29.86&\bfseries21.24  & 
\bfseries 19.54 &\bfseries45.82 & 60.97  & 11.79\\
    \bottomrule
    \end{tabular}}
\end{table*}
\endgroup

To train the model with the BIT dataset, for each sample $\boldsymbol{v}$, we sample a multi-run conversation $X_t = (X_u^1, X_a^1, \cdots, X_u^T, X_a^T)$ from all conversations associated with it, where $T \geq 1$ represents the number of turns. $a$ indicates the message from the assistant and $u$ indicates the message is from the user. The training objective is to maximize the probability of the assistant's response only
$$
\arg\max_\theta p(X_a | X_v, X_{\text{inst}}) = \prod_{t=1}^T p({\color{magenta}X_a^t} | X_u^{\leq t}, X_a^{\le t }, X_\text{inst}, X_v)
$$
Figure~\ref{fig:chat} illustrates the chat template and the training objective. We freeze the weights of the LLM and only train the fMRI encoder since we want to preserve the LLM's language modeling prior and ensure a fair comparison with baselines such as \citet{xia2024umbrae}.

\noindent\textbf{Computational Complexity} According to the analysis in Appendix~\ref{app:complexity}, our model does not introduce additional complexity compared to previous methods \cite{scotti2024mindeye2, wang2024mindbridge}.

\begin{figure}[htbp]
\centering
\begin{minipage}{0.99\columnwidth}\vspace{0mm}    \centering
\begin{tcolorbox}[colback=white,colframe=gray,left=1pt,top=1pt,bottom=1pt]
\sffamily
\footnotesize	
  \texttt{<system message>}\\
  user: $X_v$, $X_\text{inst}$, $X_1^u$ \\
  assistant: {\color{magenta}$X_1^a$}\\
user: $X_2^u$\\
  assistant: {\color{magenta}$X_2^a$}\\
  $\cdots\cdots$
\end{tcolorbox}
\end{minipage}
\caption{The chat template used during instruction tuning, illustrating two turns of conversations. Two turns of conversation are shown. Tokens highlighted in {\color{magenta}magenta} are used for next-token prediction loss computation.}
\label{fig:chat}
\end{figure}

\section{Experiments}
In this section, we first evaluate our model on various downstream tasks, demonstrating its versatile decoding capabilities. Next, we assess its generalizability to novel subjects and its adaptability to real-world applications. Finally, we analyze the functions of queries in our neuroscience-informed attention mechanism.

\subsection{Settings}

\noindent\textbf{fMRI Datasets} 
We use the widely used Natural Scenes Dataset (NSD) \cite{allen2022massive}, a large-scale dataset consisting of fMRI measurements of $8$ healthy adult subjects. During data collection, subjects viewed images from the MS-COCO dataset \cite{lin2014microsoft} and were instructed to press buttons to indicate whether they had previously seen each image.


\noindent\textbf{Downstream Datasets} The downstream dataset will be discussed within each experiment section. See examples and a short description for all datasets we will use in Appendix~\ref {app:dataset}.
Implementation details could be found in Appendix~\ref{app:impl}.

\begingroup
\sisetup{
  table-format=3.2,  
  table-align-text-pre=false,
  propagate-math-font=true,
  table-number-alignment=center,
  detect-weight=true,detect-inline-weight=math
}
\def\Uline#1{#1\llap{\uline{\phantom{#1}}}}
\begin{table*}[t]\centering
\caption{Versatile decoding. A dash $-$ means the model could not perform this task. The superscript $^\circ$ means the model is trained from scratch, in contrast to their BIT version. The CIDEr metric is scaled by a factor of 100 for consistency with Table~\ref{tab:caption} and baselines.}
\vskip 0.1in
\label{tab:versatile}
\resizebox{\linewidth}{!}{
\begin{tabular}{cc|SSS|SSS|SSS}\toprule
& & {OneLLM} & {UMBRAE} &{BrainChat} & {MindBridge$^\circ$} & {UniBrain$^\circ$} & {\makecell{\name{}}$^\circ$} &{MindBridge} & {UniBrain} & {\makecell{\name{}}} \\
\midrule
subj-agnostic &  & {\xmark}& {\xmark} & {\xmark} & {\cmark} & {\cmark} & {\cmark} &{\cmark} & {\cmark} & {\cmark} \\
\midrule
COCO-QA &Accuracy$\uparrow$ &11.09\% & 22.23\% & 39.44\% &40.19\% &38.38\% &42.09\% &\Uline{45.33\%} &42.00\% & \bfseries 48.19\% \\
\midrule
VG-QA &Accuracy$\uparrow$ & 8.76\% & 19.67\% & 21.00\% &20.84\% &21.27\% &21.68\% &23.53\% & \Uline{24.02\%} &\bfseries 24.06\% \\
\midrule
VQA-v2 &Accuracy $\uparrow$& 33.68\% & \Uline{51.23}\% & 40.02\% &43.25\% &46.04\% &44.13\% &47.91\% &48.58\% &\bfseries 52.14\% \\
\midrule
A-OKVQA &Accuracy $\uparrow$&25.23\%& 43.24\% &20.52\% &22.12\% &19.47\% &29.20\% & \Uline{50.44\%} &43.36\% &\bfseries 52.21\% \\
\midrule
ST-VQA &ANLS $\uparrow$& 5.74\%& 5.46\% &9.58\% &10.20\% &7.01\% &\Uline{12.76}\% &11.64\% &8.76\% &\bfseries 12.92\% \\
\midrule
OK-VQA &Accuracy $\uparrow$&22.98\% & 10.35\% &17.22\%&27.63\% &18.63\% &27.70\% &32.13\% &\Uline{32.30\%} & \bfseries 33.33\% \\
\midrule
\multirow{2}{*}{TallyQA} &Accuracy $\uparrow$& 8.34\% & 44.10\% & 43.22\%&43.49\% &44.83\% &43.75\% &49.46\% & \Uline{53.77\%} &\bfseries 54.76\% \\
&RMSE $\downarrow$&7.45 & 3.94 & 1.90 &2.03 &1.83 &2.04 &1.86 &\bfseries 1.67 &\Uline{1.76} \\
\midrule
\multirow{6}{*}{Paragraph Caption} &BLEU-1$\uparrow$ &0.26 & \bfseries 29.82 & 22.21&21.82 &25.69 &26.49 &25.69 &28.28 & \Uline{29.43} \\
&BLEU-2 $\uparrow$ &0.08 & 14.26 &10.23 &10.47 &12.62 &12.48 &13.00 &\Uline{15.47} & \bfseries 15.78 \\
&BLEU-3$\uparrow$ &0.03 & 6.52 &6.38 &5.58 &6.70 &6.43 &7.10 &\Uline{8.90} & \bfseries 9.14 \\
&BLEU-4$\uparrow$ &0.01 & 2.95 &2.12 &3.14 &3.81 &3.63 &4.22 &\bfseries5.60 & \Uline{5.51} \\
&METEOR $\uparrow$&2.36 &12.60 &9.10 &10.95 &11.13 &10.71 &11.39 &\bfseries13.50 & \Uline{13.18} \\
&CIDEr $\uparrow$&0.00 & 7.39 &6.02& \Uline{7.50} &3.92 &2.44 & 3.55 &1.82 & \bfseries 7.80 \\
\midrule
\multirow{8}{*}{VQA-E} &Accuracy $\uparrow$ &19.60\%& 47.84\% &46.20\% &45.40\% &44.42\% &44.55\% &\Uline{48.48\%} &48.39\% &\bfseries 50.95\% \\
&BLEU-1 $\uparrow$&17.32& 29.83 & 35.99 &35.63 &35.30 &35.08 &36.18 &\Uline{37.26} &\bfseries 37.70 \\
&BLEU-2 $\uparrow$&7.44& 14.76 & 18.33  &18.27 &18.04 &17.82 &19.38 &\Uline{20.41} &\bfseries 20.56 \\
&BLEU-3 $\uparrow$&3.62& 8.17 & 10.01 &10.32 &10.20 &10.05 &11.30 &\Uline{12.25} &\bfseries 12.34 \\
&BLEU-4 $\uparrow$&1.82& 4.87 &6.60 &6.27 &6.14 &6.00 &7.00 &\Uline{7.83} &\bfseries 7.92 \\
&CIDEr $\uparrow$&19.32& 63.26 &78.33 &79.05 &77.31 &76.80 &86.62 &\Uline{92.09} & \bfseries 93.60 \\
&METEOR $\uparrow$&6.69 & 12.25 &13.64 &14.13 &13.89 &13.96 &14.81 &\Uline{15.51} &\bfseries 15.62 \\
&ROUGE $\uparrow$&16.84& 28.38 & 32.82&33.78 &33.25 &33.11 &34.56 &\Uline{35.87} &\bfseries 35.88 \\
\midrule
\multirow{8}{*}{FSVQA} &VQA Acc. $\uparrow$& 31.44\% & 40.67\% &36.30\% &42.00\% &37.05\% &42.53\% &\Uline{45.95\%} &44.58\% &\bfseries 48.03\% \\
&FSVQA Acc. $\uparrow$& 21.02\% & 0.00\% & 30.22\% &37.40\% &32.30\% &38.50\% &\Uline{40.97\%} &37.87\% &\bfseries 43.00\% \\
&BLEU-1 $\uparrow$& 37.42 & 23.11 & 83.99&85.68 &83.84 &85.88 &\Uline{86.52} &85.10 &\bfseries 87.10 \\
&BLEU-2 $\uparrow$ & 31.72 & 5.86&78.50 &81.27 &78.81 &81.62 &\Uline{82.28} &80.01 &\bfseries 83.03 \\
&BLEU-3 $\uparrow$& 26.95 & 2.10&73.00 &77.10 &73.97 &77.62 &\Uline{78.34} &75.49 &\bfseries 79.27 \\
&BLEU-4 $\uparrow$& 22.48 & 1.04 &69.73 &72.89 &68.91 &73.56 &\Uline{74.35} &70.73 &\bfseries 75.50 \\
&METEOR $\uparrow$& 26.35 & 8.93 &44.76 &47.59 &45.94 &47.96 &\Uline{48.63} &46.89 &\bfseries 49.05 \\
&CIDEr $\uparrow$& 312.75 & 4.07 &600.00 &636.40 &609.00 &646.26 &\Uline{657.02} &628.83 &\bfseries 666.26 \\
\midrule
\multirow{8}{*}{Previous Caption} & BLEU-1 $\uparrow$ & 41.86  & {$-$} & 21.19 & 21.17 & 24.84 & \Uline{44.52} & 42.45 & 43.01 & \bfseries 47.20 \\
&BLEU-2 $\uparrow$ & 19.44 & {$-$} & 8.00 & 7.57 & 9.70 & \Uline{22.46} & 20.04 & 20.03 & \bfseries 25.16\\
&BLEU-3 $\uparrow$ & 9.25 & {$-$} & 1.98 & 2.85 & 3.40 & \Uline{10.39} & 9.61 & 9.19 & \bfseries 12.95\\
&BLEU-4 $\uparrow$ & 3.67 & {$-$} & 1.02& 1.28 & 1.46 & \Uline{5.45} &  5.31 & 4.58 & \bfseries 7.49 \\
&METEOR $\uparrow$ & 10.14 & {$-$} & 6.55& 6.46 & 7.20 & \Uline{11.00} & 10.83 & 10.81 & \bfseries 11.96 \\
&ROUGE $\uparrow$ & 30.19 & {$-$} & 21.23& 20.88 & 23.04 & \Uline{33.20} &32.38 & 31.99 & \bfseries 34.58 \\
&CIDEr $\uparrow$ & 6.65 & {$-$} & 9.21& 8.83 & \Uline{11.73} & 9.39 & 7.89 & 7.53 & \bfseries 16.02 \\
&SPICE $\uparrow$ & 2.49 & {$-$} & 2.44& 2.56 & 2.78 & \Uline{3.07} & 2.80 & 2.92 & \bfseries 3.93 \\
\bottomrule
\end{tabular}}
\end{table*}
\endgroup

\subsection{Brain Captioning}
To evaluate the model's performance on downstream tasks, we start with the widely used brain captioning benchmark \cite{xia2024umbrae}. The task, built upon COCO Caption \cite{chen2015microsoft} requires the model to predict captions of given images as fMRI stimuli.

\noindent\textbf{Baselines}
The following baselines are considered in this experiment: SDRecon \cite{takagi2023high}, UniBrain \cite{mai2023unibrain}, and BrainCap \cite{ferrante2023brain} employs a linear regression, mapping the fMRI to the inputs of an image caption model \cite{li2023blip}. OneLLM \cite{han2024onellm} is a multimodal large language models that align $8$ modalities (including fMRI) with language all in one model. For fair and efficient comparison, we only finetune the encoder, given that we freeze the LLM in our method as well. UMBRAE learns an encoder that maps fMRIs to images through an encoder similar to the MLP mixer \cite{tolstikhin2021mlp}. BrainChat \cite{huang2024brainchat} segments the flattened voxels into 16 patches and employs a transformer to decode text conditioned on the patches.
It is worth noting that all of these baselines require subject-specific layers or parameters. In contrast, our model is subject-agnostic, thus with the potential to generalize on novel subjects.

\noindent\textbf{Metric} Following previous works, we use five standard metrics for text generation: BLEU-$k$ \cite{papineni2002bleu}, ROUGE-L \cite{lin2004rouge}, CIDEr \cite{vedantam2015cider}, SPICE \cite{anderson2016spice}, METEOR \cite{banerjee2005meteor}.

Table \ref{tab:caption} shows that our model outperforms baselines in terms of most metrics, with an average improvement of $3.32\%$, even if our model does not have any subject-specific layers. We argue that this is attributed to both the novel architecture design and the introduction of BIT, which will be evident in the next experiment.

\begingroup
\sisetup{
  table-format=3.2,  
  table-align-text-pre=false,
  table-number-alignment=center,
detect-weight=true,detect-inline-weight=math
}
\begin{table}[t]
\vspace{-1em}
\centering
\caption{Model generalization, compared with subject-agnostic model. We train the models on subject $1-7$ and evaluate on subject $8$, which is the held-out subject.}
\label{tab:subj8}
\vskip 0.1in

\resizebox{\linewidth}{!}{
\begin{tabular}{ccSSS}\toprule
& & {MindBridge} & {UniBrain} & {\name{}} \\
\midrule
COCO-QA &Accuracy $\uparrow$&35.88 &24.95 &\bfseries 38.96 \\
\midrule
VG-QA &Accuracy $\uparrow$&\bfseries 20.56 &16.23 & 19.29 \\
\midrule
VQA-v2 &Accuracy $\uparrow$&42.80 &40.16 &\bfseries 46.04 \\
\midrule
A-OKVQA &Accuracy $\uparrow$&44.55 &28.71 &\bfseries 45.54 \\
\midrule
ST-VQA &ANLS $\uparrow$&9.33 &9.30 &\bfseries 13.63 \\
\midrule
OK-VQA &Accuracy $\uparrow$&21.94 &17.09 &\bfseries 25.44 \\
\midrule
\multirow{2}{*}{TallyQA} &Accuracy $\uparrow$&38.92 &32.51 &\bfseries 42.29 \\
&RMSE $\downarrow$&2.12 &\bfseries 2.02 & 2.11 \\
\midrule
\multirow{8}{*}{COCO-Caption} &BLEU-1 $\uparrow$&39.84 &41.90 &\bfseries 44.60 \\
&BLEU-2 $\uparrow$&19.55 &19.67 &\bfseries 24.04 \\
&BLEU-3 $\uparrow$&9.29 &8.89 &\bfseries 12.79 \\
&BLEU-4 $\uparrow$&5.24 &4.33 &\bfseries 7.52\\
&METEOR $\uparrow$&10.39 &10.80 &\bfseries 11.16 \\
&ROUGE $\uparrow$&31.10 &31.54 &\bfseries 33.47 \\
&CIDEr $\uparrow$& 8.70 &6.40 & \bfseries 13.22 \\
&SPICE $\uparrow$&2.67 &2.39 &\bfseries 3.82 \\
\midrule
\multirow{6}{*}{Paragraph Caption} &BLEU-1 $\uparrow$&23.18 &21.73 &\bfseries 26.07 \\
&BLEU-2 $\uparrow$&10.71 &8.94 &\bfseries 12.64 \\
&BLEU-3 $\uparrow$&4.61 &3.72 &\bfseries 6.27 \\
&BLEU-4 $\uparrow$&2.22 &1.92 &\bfseries 3.46 \\
&METEOR $\uparrow$&9.99 &9.47 &\bfseries 11.18 \\
&CIDEr $\uparrow$&0.71 &1.56 &\bfseries 5.61 \\
\midrule
\multirow{8}{*}{VQA-E} &Accuracy $\uparrow$&41.78 &38.53 &\bfseries 46.19 \\
&BLEU-1 $\uparrow$&32.54 &32.86 &\bfseries 34.54 \\
&BLEU-2 $\uparrow$&16.13 &15.48 &\bfseries 17.54 \\
&BLEU-3 $\uparrow$&8.82 &7.98 &\bfseries 9.84 \\
&BLEU-4 $\uparrow$&5.16 &4.42 &\bfseries 5.81 \\
&CIDEr $\uparrow$&68.13 &58.79 &\bfseries 73.27 \\
&METEOR $\uparrow$&12.74 &12.26 &\bfseries 13.56 \\
&ROUGE $\uparrow$&30.63 &29.38 &\bfseries 32.36 \\
\midrule
\multirow{8}{*}{FSVQA} &VQA Acc. $\uparrow$&42.33 &37.92 &\bfseries 43.92 \\
&FSVQA Acc. $\uparrow$&37.16 &30.83 &\bfseries 39.07 \\
&BLEU-1 $\uparrow$&75.81 &82.94 &\bfseries 86.19 \\
&BLEU-2 $\uparrow$&71.03 &77.24 &\bfseries 81.83 \\
&BLEU-3 $\uparrow$&66.40 &71.94 &\bfseries 77.77 \\
&BLEU-4 $\uparrow$&61.55 &66.54 &\bfseries 73.58 \\
&METEOR $\uparrow$&45.84 &45.24 &\bfseries 47.94 \\
&CIDEr $\uparrow$&428.39 &587.78 &\bfseries 646.30 \\
\bottomrule
\end{tabular}
}
\vspace{-1em}
\end{table}
\endgroup

\subsection{Versatile Decoding}
The purpose of experiments in this section is two-fold: 1) To investigate the impact of our model design and the introduction of BIT on performance improvement. 2) To evaluate the versatility of the model, i.e., its performance on various downstream tasks.

\noindent\textbf{Baselines} Besides baselines that could be adapted to this experiment from the previous one, we further consider the following subject-agnostic models as baselines.
1) MindBridge \cite{wang2024mindbridge} flattens the voxels and adaptively adjusts the padding and stride to pool the voxels into a fixed dimension. The original implementation of MindBridge has subject-specific parameters. However, since those parameters are of the same size, we make them shared across subjects and thus make the model subject-agnostic.
2) UniBrain \cite{wang2024unibrain} samples voxels into a fixed number of groups and employs a transformer where groups are treated as tokens. This UniBrain is unrelated to the UniBrain in the previous section; they just share the same name.

\noindent\textbf{Datasets \& Metric}
We use the test split of all QA \& caption datasets in the BIT dataset. We strictly adhere to the official metrics on all datasets. In summary, for sentence generation, we use BLEU-$k$~\cite{papineni2002bleu}, ROUGE-L~\cite{lin2004rouge}, CIDEr~\cite{vedantam2015cider}, SPICE~\cite{anderson2016spice}, METEOR~\cite{banerjee2005meteor}. For QA-related tasks, we use VQA accuracy~\cite{antol2015vqa} as well as special metrics proposed in the original paper (e.g. ANLS for ST-VQA~\cite{biten2019scene}).

The results on subject 1 are shown in Table~\ref{tab:versatile}. Our model outperforms baselines, with an average improvement of $12.0\%$. Further, by comparing instruction tuning and from-scratch models, we find that instruction tuning has a significant positive effect, with an average improvement of $28.0\%$. The results remain stable across different random seeds; for instance, according to our observations, the BLEU-1 score for paragraph captioning exhibits a maximum of $\pm 0.3$ variance. We report experiments on other subjects in Appendix~\ref{app:breakdown}.

\begingroup
\sisetup{
  table-format=2.2,  
  table-align-text-pre=false,
  table-number-alignment=center,
detect-weight=true,detect-inline-weight=math
}
\begin{table}[htbp]
    \caption{Model adaptation to new tasks. \textit{sentiment understanding} and \textit{utility/affordance} are sub-datasets from TDIUC that are particularly relevant to BCI applications.}
    \label{tab:new_task}
\vskip 0.1in
    \centering
    \resizebox{\linewidth}{!}{
    \begin{tabular}{cSSSS}
    \toprule
    \multirow{2}{*}{Method} &  \multicolumn{2}{c}{Overall} & {Sentiment Understanding} & {Utility/Affordance} \\
    \cmidrule{2-3} \cmidrule{4-4} \cmidrule{5-5}
    & {A-MPT} & {H-MPT} & {Accuracy} & {Accuracy} \\
    \midrule
    \name{}$^\circ$ & 41.09\% & 19.38\% & 70.00\% & 0.00\%\\
    \midrule
         MindBridge & 49.77\% & 39.88\% & 80.00\% & 14.29\%\\
         UniBrain & 51.50\% & 36.76\% & 80.00\% & 28.57\%\\
         \name{}& \bfseries 54.08\% &  \bfseries 45.43\% & \bfseries 80.77\% & \bfseries 50.00\% \\
    \bottomrule
    \end{tabular}
    }
    \vspace{-1em}
\end{table}
\endgroup

\subsection{Unseen Subject Generalization}
Our neuroscience-informed, subject-agnostic design enhances generalization to novel subjects, a crucial factor in real-world applications where training a model for each individual is impractical. To evaluate it, we perform instruction tuning on $7$ out of the $8$ subjects in the natural scene dataset \cite{allen2022massive}, and evaluate generalization on the held-out subject. Table~\ref{tab:subj8} shows our model outperforms two other subject-agnostic baselines in most cases, with an average improvement of $24.5\%$ compared to the second-best model. We anticipate further performance gains through the application of domain adaptation techniques \cite{xiao2023spa,ganin2015unsupervised,ganin2016domain,gong2012geodesic}, which we leave as a direction for future work.

\subsection{Adapting to New Tasks}
It is common that users want to adapt the \name{} to their own specific use cases. To this end, we aim to assess our model's adaptability to new tasks.

\noindent\textbf{Dataset \& Metrics} We use TDIUC \cite{kafle2017analysis}, a QA dataset consisting of $12$ types of questions, as a benchmark to evaluate the model's various capabilities comprehensively. Additionally, we further select $2$ task types-\textit{sentiment understanding} and \textit{utility/affordance} tasks, that are particularly relevant to BCI applications as sub-datasets. The \textit{utility/affordance} task, for instance, enables the model to identify useful objects in a given scene and autonomously decide whether to utilize them. Following their paper, we compute the accuracy of each type and report the arithmetic mean-per-type (A-MPT) and the harmonic mean-per-type (H-MPT). For the $2$ selected types, we report the accuracy respectively. 

Table~\ref{tab:new_task} shows our model achieves balanced (high harmonic mean) and consistently improved performances with an average of $13.5\%$. We could also observe the performance benefits from BIT, with $25.0\%$ absolute improvement.

\subsection{Performance scale with the number of subjects}

We conducted experiments to evaluate how model performance scales with the number of subjects and report performances on the COCO caption task. We examined both the in-distribution (seen subjects) setting in Figure~\ref{tab:in} and the out-of-distribution (held-out subjects) setting in Figure~\ref{tab:out}. Our results show significant performance improvements as the number of training subjects increases, demonstrating that the model benefits from exposure to more subjects during pre-training.

\begingroup
\sisetup{
  table-format=3.2,  
  table-align-text-pre=false,
  table-number-alignment=center,
detect-weight=true,detect-inline-weight=math
}
\begin{table}[!htp]\centering
\caption{Performance scaling with number of subjects (in-distribution). Evaluation is performed on Subject 1, who is included during training.}\label{tab:in}
\vskip 0.1in
\resizebox{\linewidth}{!}{
\begin{tabular}{cSSSSSSSSS}\toprule
& \multicolumn{8}{c}{COCO Caption}  \\
\makecell{number\\of\\subjects} &{{BLEU-1} $\uparrow$} & {BLEU-2 $\uparrow$} & {BLEU-3 $\uparrow$} & {{BLEU-4} $\uparrow$} &{METEOR $\uparrow$}&{ROUGE $\uparrow$}& {CIDEr $\uparrow$}&{SPICE $\uparrow$}\\
\midrule
1 &58.05 &37.95 &24.40 &16.14 &16.62 &42.03 &43.04 &8.81 \\
3 &56.81 &37.16 &24.12 &16.30 &16.44 &41.93 &40.94 &9.12 \\
5 &58.17 &38.65 &25.49 &17.30 &16.97 &42.36 &46.54 &9.19 \\
7 &58.52 &39.20 &25.95 &17.51 &17.12 &42.87 &47.45 &9.48 \\
\bottomrule
\end{tabular}
}
\end{table}

\begin{table}[!htp]\centering
\caption{Performance scaling with number of subjects (out-of-distribution). Evaluation is performed on Subject 8, who is held out during training.}\label{tab:out}
\vskip 0.1in
\resizebox{\linewidth}{!}{
\begin{tabular}{cSSSSSSSSS}\toprule
\makecell{number\\of\\subjects} &{{BLEU-1} $\uparrow$} & {BLEU-2 $\uparrow$} & {BLEU-3 $\uparrow$} & {{BLEU-4} $\uparrow$} &{METEOR $\uparrow$}&{ROUGE $\uparrow$}& {CIDEr $\uparrow$}&{SPICE $\uparrow$}\\
\midrule
1 &42.40 &19.51 &8.61 &4.19 &9.48 &30.68 &5.41 &1.70 \\
3 &43.95 &22.45 &10.37 &5.53 &10.37 &32.49 &5.79 &2.59 \\
5 &45.00 &22.61 &10.64 &5.60 &10.53 &32.53 &6.22 &2.79 \\
7 &47.30 &25.35 &13.61 &8.15 &11.40 &34.64 &6.41 &3.61 \\
\bottomrule
\end{tabular}
} 
\end{table}

\endgroup

\subsection{Ablation Study}
\begin{figure}[h]
    \centering
    \includegraphics[width=\linewidth]{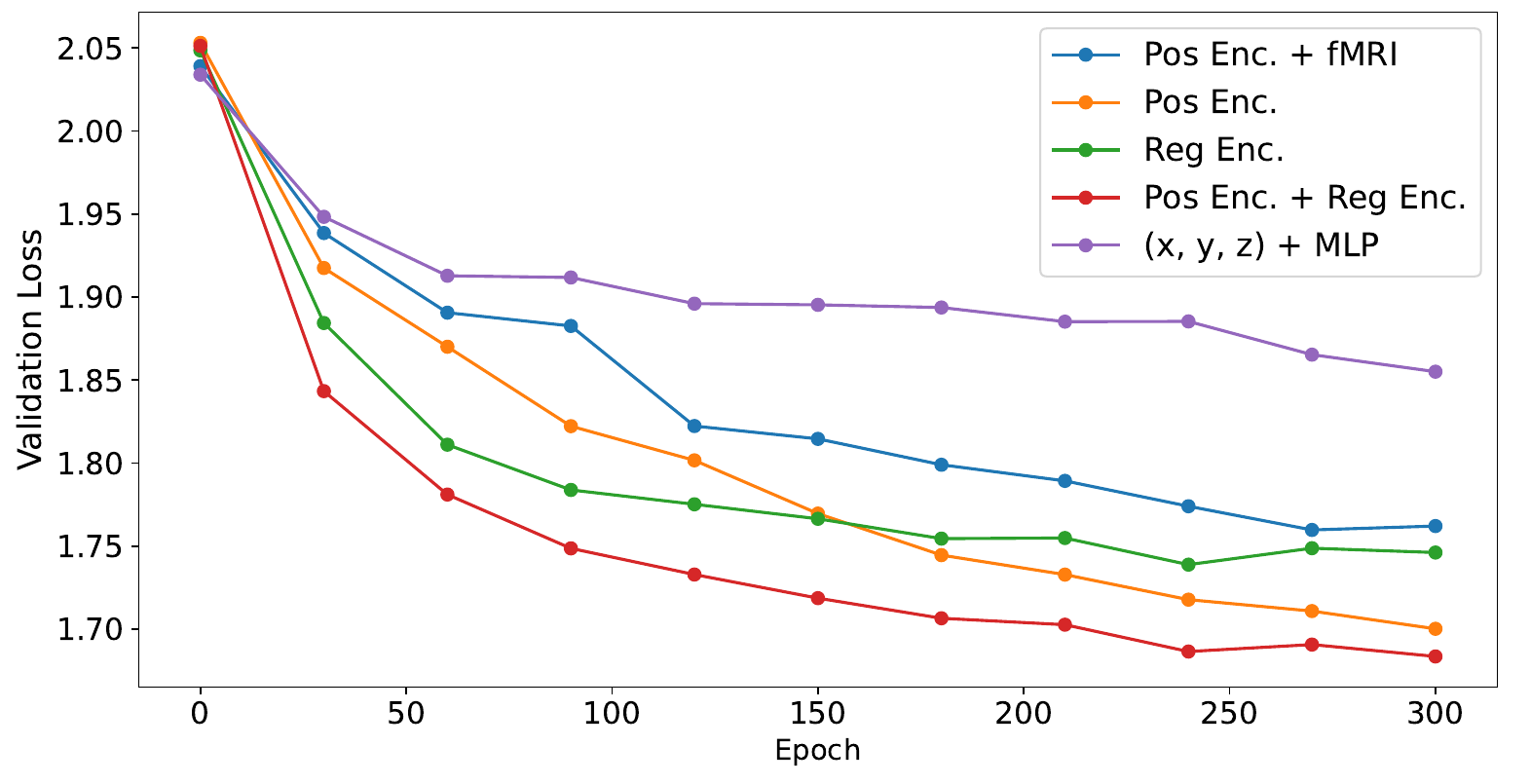}
    \vspace{-2em}
    \caption{Ablation study of the key embedding design. Pos Enc. stands for positional encoding. Reg Enc. stands for multiple region encodings. \textit{(x, y, z)+MLP} means we employ an MLP to map the coordinates to the embeddings instead of positional encoding.}
    \label{fig:ablation}
\end{figure}

We conduct ablation studies on the design of key embeddings in the neuroscience-informed attention module in Figure~\ref{fig:ablation}. The results strongly validate our design. The vanilla cross attention (\textit{Pos Enc.+fMRI}) leads to poor performance, while removing fMRI values from the key embeddings (\textit{Pos Enc.}) yields a significant improvement. Replacing positional encoding with region encodings (\textit{Reg. Enc.}) accelerates convergence in the early stages since it is grounded by neuroscientific principles. However, it is eventually outperformed by \textit{Pos Enc.} due to the lack of finer-grained information. Combining the positional encoding and region encodings (\textit{Pos Enc.+Reg Enc.}) achieves the best model design. In addition, replacing positional encoding with an MLP that maps coordinates to embeddings results in poor performance (\textit{(x,y,z)+MLP}), which indicates the amount of high-frequency spatial information in fMRI signals. Additional ablation results on evaluation metrics are provided in Appendix~\ref{app:ablation}.

\subsection{Visualizations and Interpretations}
\label{sec:vis}

\begin{figure}
    \centering
    \includegraphics[width=\linewidth]{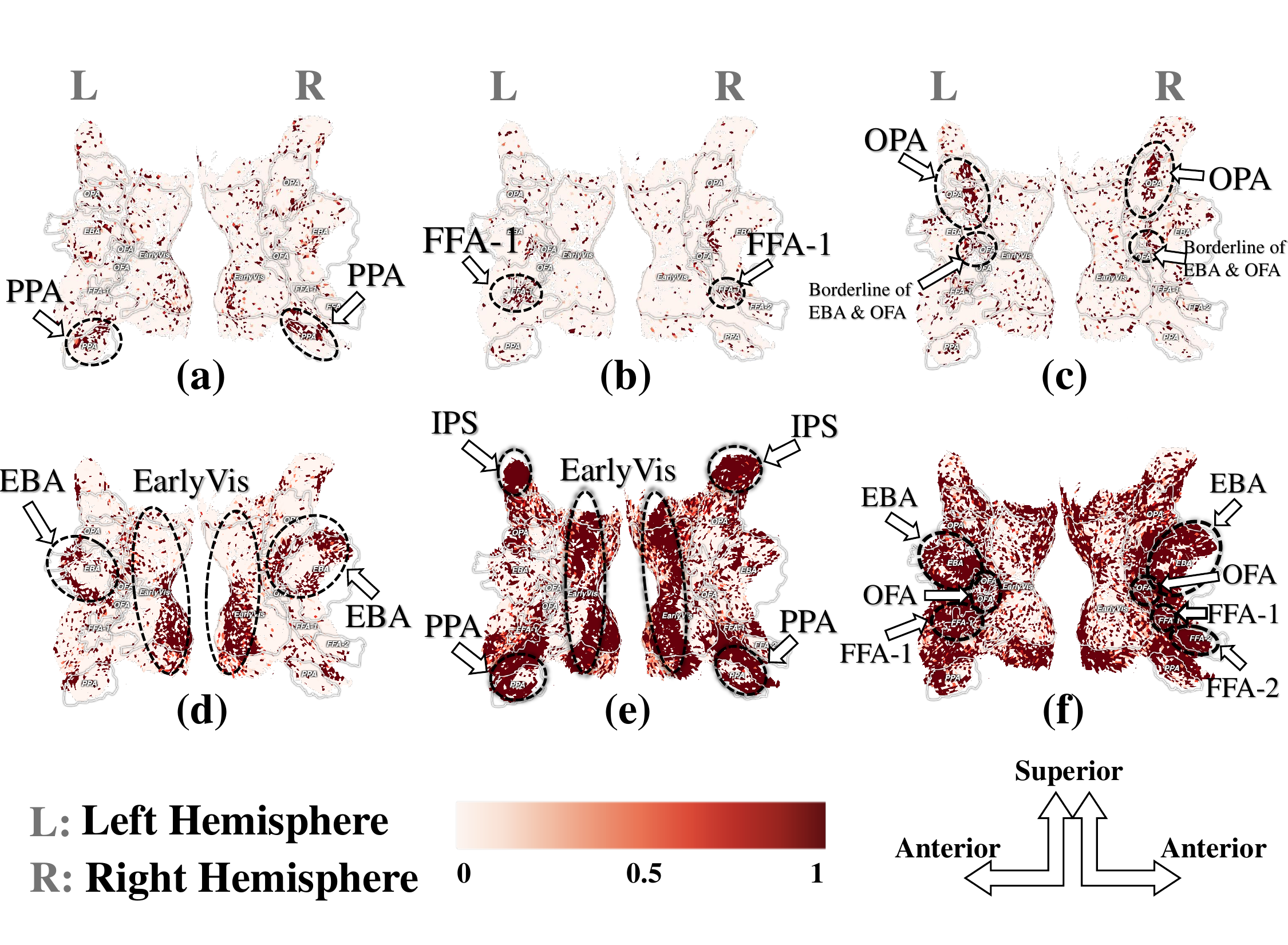}
    \caption{Each subfigure corresponds to a specific query of subject 1. We visualize the attention map between that query token and all voxels. Here we randomly select 6 query tokens, each of which exhibits a distinct spatial focus.}
    \label{fig:query}
\end{figure}

Unlike previous deep learning models \cite{scotti2024mindeye2,mai2023unibrain}, our model allows interpretations by investigating how \textit{queries} work in the neuroscience-informed attention layer. We inspect the attention weights between queries and voxels in Figure~\ref{fig:query}. 

We found that some queries primarily focus on processing single brain regions, such as Parahippocampal Place Area (PPA) (Figure~\ref{fig:query}a) and Fusiform Face Area (FFA) (\ref{fig:query}b). As previous research has shown, PPA is related to conceptual association, semantic processing and environmental memory \cite{epstein1999parahippocampal,kohler2002differential,bar2008scenes,epstein2010reliable} and FFA is known for its critical role in expertise recognition, social cognition and identity memory \cite{schultz2003role,tsantani2021ffa,xu2005revisiting}. Both are important brain regions for the conceptualization of visual information and are responsible for the interaction between real-time stimulus and past memory \cite{brewer1998making,ranganath2004category,golarai2007differential}. 

Moreover, there are some queries that attend to multiple brain regions, revealing the information transmission between low- and high-level brain regions. For instance, interactions between early visual areas and higher-level regions like PPA and IntraParietal Sulcus (IPS) (Figure~\ref{fig:query}e), revealing a potential pattern for human attention-guided actions \cite{tunik2007beyond,connolly2016coding}. Additionally, queries are also found responsible for communications between high-level brain regions (Figure \ref{fig:query}c,\ref{fig:query}f). Together, these findings indicate that the learnable queries may reflect the dynamics of human brain activities in the visual task, from seeing and thinking about the image to pressing the button for the visual recall task in NSD \cite{allen2022massive}.

We provide attention map visualizations of subjects other than subject 1 in Appendix~\ref{app:brainmap_other}. We also visualize latent embeddings during the model forward pass in Appendix~\ref{app:latent}. Qualitative analysis of model responses could be found in Appendix~\ref{app:qual}.

\section{Conclusion}
In this work, we propose \name{}, a subject-agnostic and versatile fMRI-to-text decoding model. Our neuroscience-informed attention mechanism in the fMRI encoder ensures subject generalization, while brain instruction tuning enhances versatility. Comprehensive benchmarks demonstrate our model's state-of-the-art performance, and visualizations offer insights into the patterns it leverages. We envision that these advancements will contribute to medical treatments, neuroscience research, and brain-computer interfaces in the future. \noindent\textbf{Limitations:} This work focuses on static fMRI, without incorporating temporal dynamics. Future research could explore integrating temporal information and investigating its relationship with other modalities such as videos.

\section*{Acknowledgments}
We are grateful for funding and computing resource support from
Amazon Research, Snap Research, National Science Foundation (NSF) IIS Div Of Information \& Intelligent Systems 2403317, Yale AI Engineering Research Seed Grants, Yale Office of the Provost, and Yale Wu Tsai Institute.

\section*{Impact Statement}
This paper presents work whose goal is to advance the field of Machine Learning. There are many potential societal consequences of our work, none of which we feel must be specifically highlighted here.

\bibliography{example_paper}

\begin{thebibliography}{98}
\providecommand{\natexlab}[1]{#1}
\providecommand{\url}[1]{\texttt{#1}}
\expandafter\ifx\csname urlstyle\endcsname\relax
  \providecommand{\doi}[1]{doi: #1}\else
  \providecommand{\doi}{doi: \begingroup \urlstyle{rm}\Url}\fi

\bibitem[Acharya et~al.(2019)Acharya, Kafle, and Kanan]{acharya2019tallyqa}
Acharya, M., Kafle, K., and Kanan, C.
\newblock Tallyqa: Answering complex counting questions.
\newblock In \emph{Proceedings of the AAAI conference on artificial intelligence}, volume~33, pp.\  8076--8084, 2019.

\bibitem[Alayrac et~al.(2022)Alayrac, Donahue, Luc, Miech, Barr, Hasson, Lenc, Mensch, Millican, Reynolds, et~al.]{alayrac2022flamingo}
Alayrac, J.-B., Donahue, J., Luc, P., Miech, A., Barr, I., Hasson, Y., Lenc, K., Mensch, A., Millican, K., Reynolds, M., et~al.
\newblock Flamingo: a visual language model for few-shot learning.
\newblock \emph{Advances in neural information processing systems}, 35:\penalty0 23716--23736, 2022.

\bibitem[Allen et~al.(2022)Allen, St-Yves, Wu, Breedlove, Prince, Dowdle, Nau, Caron, Pestilli, Charest, et~al.]{allen2022massive}
Allen, E.~J., St-Yves, G., Wu, Y., Breedlove, J.~L., Prince, J.~S., Dowdle, L.~T., Nau, M., Caron, B., Pestilli, F., Charest, I., et~al.
\newblock A massive 7t {fMRI} dataset to bridge cognitive neuroscience and artificial intelligence.
\newblock \emph{Nature neuroscience}, 25\penalty0 (1):\penalty0 116--126, 2022.

\bibitem[Anderson et~al.(2016)Anderson, Fernando, Johnson, and Gould]{anderson2016spice}
Anderson, P., Fernando, B., Johnson, M., and Gould, S.
\newblock Spice: Semantic propositional image caption evaluation.
\newblock In \emph{Computer Vision--ECCV 2016: 14th European Conference, Amsterdam, The Netherlands, October 11-14, 2016, Proceedings, Part V 14}, pp.\  382--398. Springer, 2016.

\bibitem[Antol et~al.(2015)Antol, Agrawal, Lu, Mitchell, Batra, Zitnick, and Parikh]{antol2015vqa}
Antol, S., Agrawal, A., Lu, J., Mitchell, M., Batra, D., Zitnick, C.~L., and Parikh, D.
\newblock Vqa: Visual question answering.
\newblock In \emph{Proceedings of the IEEE international conference on computer vision}, pp.\  2425--2433, 2015.

\bibitem[Banerjee \& Lavie(2005)Banerjee and Lavie]{banerjee2005meteor}
Banerjee, S. and Lavie, A.
\newblock Meteor: An automatic metric for mt evaluation with improved correlation with human judgments.
\newblock In \emph{Proceedings of the acl workshop on intrinsic and extrinsic evaluation measures for machine translation and/or summarization}, pp.\  65--72, 2005.

\bibitem[Bar et~al.(2008)Bar, Aminoff, and Schacter]{bar2008scenes}
Bar, M., Aminoff, E., and Schacter, D.~L.
\newblock Scenes unseen: the parahippocampal cortex intrinsically subserves contextual associations, not scenes or places per se.
\newblock \emph{Journal of Neuroscience}, 28\penalty0 (34):\penalty0 8539--8544, 2008.

\bibitem[Bernal et~al.(2022)Bernal, P{\'e}rez, Beltr{\'a}n, P{\'e}rez, and Celdr{\'a}n]{bernal2022brain}
Bernal, S.~L., P{\'e}rez, M.~Q., Beltr{\'a}n, E. T.~M., P{\'e}rez, G.~M., and Celdr{\'a}n, A.~H.
\newblock When brain-computer interfaces meet the metaverse: Landscape, demonstrator, trends, challenges, and concerns.
\newblock \emph{arXiv preprint arXiv:2212.03169}, 2022.

\bibitem[Biten et~al.(2019)Biten, Tito, Mafla, Gomez, Rusinol, Valveny, Jawahar, and Karatzas]{biten2019scene}
Biten, A.~F., Tito, R., Mafla, A., Gomez, L., Rusinol, M., Valveny, E., Jawahar, C., and Karatzas, D.
\newblock Scene text visual question answering.
\newblock In \emph{Proceedings of the IEEE/CVF international conference on computer vision}, pp.\  4291--4301, 2019.

\bibitem[Brewer et~al.(1998)Brewer, Zhao, Desmond, Glover, and Gabrieli]{brewer1998making}
Brewer, J.~B., Zhao, Z., Desmond, J.~E., Glover, G.~H., and Gabrieli, J.~D.
\newblock Making memories: brain activity that predicts how well visual experience will be remembered.
\newblock \emph{Science}, 281\penalty0 (5380):\penalty0 1185--1187, 1998.

\bibitem[Card et~al.(2024)Card, Wairagkar, Iacobacci, Hou, Singer-Clark, Willett, Kunz, Fan, Vahdati~Nia, Deo, et~al.]{card2024accurate}
Card, N.~S., Wairagkar, M., Iacobacci, C., Hou, X., Singer-Clark, T., Willett, F.~R., Kunz, E.~M., Fan, C., Vahdati~Nia, M., Deo, D.~R., et~al.
\newblock An accurate and rapidly calibrating speech neuroprosthesis.
\newblock \emph{New England Journal of Medicine}, 391\penalty0 (7):\penalty0 609--618, 2024.

\bibitem[Chen et~al.(2023{\natexlab{a}})Chen, Qi, Wang, and Pan]{chen2023mindgpt}
Chen, J., Qi, Y., Wang, Y., and Pan, G.
\newblock Mindgpt: Interpreting what you see with non-invasive brain recordings.
\newblock \emph{arXiv preprint arXiv:2309.15729}, 2023{\natexlab{a}}.

\bibitem[Chen et~al.(2023{\natexlab{b}})Chen, Zhang, Zeng, Zhang, Zhu, and Zhao]{chen2023shikra}
Chen, K., Zhang, Z., Zeng, W., Zhang, R., Zhu, F., and Zhao, R.
\newblock Shikra: Unleashing multimodal llm's referential dialogue magic.
\newblock \emph{arXiv preprint arXiv:2306.15195}, 2023{\natexlab{b}}.

\bibitem[Chen et~al.(2015)Chen, Fang, Lin, Vedantam, Gupta, Doll{\'a}r, and Zitnick]{chen2015microsoft}
Chen, X., Fang, H., Lin, T.-Y., Vedantam, R., Gupta, S., Doll{\'a}r, P., and Zitnick, C.~L.
\newblock Microsoft coco captions: Data collection and evaluation server.
\newblock \emph{arXiv preprint arXiv:1504.00325}, 2015.

\bibitem[Chen et~al.(2024)Chen, Du, Liu, Wang, and He]{chen2024open}
Chen, X., Du, C., Liu, C., Wang, Y., and He, H.
\newblock Open-vocabulary auditory neural decoding using fmri-prompted llm.
\newblock \emph{arXiv preprint arXiv:2405.07840}, 2024.

\bibitem[Chen et~al.(2023{\natexlab{c}})Chen, Qing, Xiang, Yue, and Zhou]{chen2023seeingbrainconditionaldiffusion}
Chen, Z., Qing, J., Xiang, T., Yue, W.~L., and Zhou, J.~H.
\newblock Seeing beyond the brain: Conditional diffusion model with sparse masked modeling for vision decoding, 2023{\natexlab{c}}.
\newblock URL \url{https://arxiv.org/abs/2211.06956}.

\bibitem[Chen et~al.(2023{\natexlab{d}})Chen, Qing, and Zhou]{chen2023cinematicmindscapeshighqualityvideo}
Chen, Z., Qing, J., and Zhou, J.~H.
\newblock Cinematic mindscapes: High-quality video reconstruction from brain activity, 2023{\natexlab{d}}.
\newblock URL \url{https://arxiv.org/abs/2305.11675}.

\bibitem[Cheng et~al.(2024)Cheng, Leng, Zhang, Xin, Li, Chen, Zhu, Zhang, Luo, Zhao, et~al.]{cheng2024videollama}
Cheng, Z., Leng, S., Zhang, H., Xin, Y., Li, X., Chen, G., Zhu, Y., Zhang, W., Luo, Z., Zhao, D., et~al.
\newblock Videollama 2: Advancing spatial-temporal modeling and audio understanding in video-llms.
\newblock \emph{arXiv preprint arXiv:2406.07476}, 2024.

\bibitem[Connolly et~al.(2016)Connolly, Kentridge, and Cavina-Pratesi]{connolly2016coding}
Connolly, J.~D., Kentridge, R.~W., and Cavina-Pratesi, C.
\newblock Coding of attention across the human intraparietal sulcus.
\newblock \emph{Experimental brain research}, 234:\penalty0 917--930, 2016.

\bibitem[Denk et~al.(2023)Denk, Takagi, Matsuyama, Agostinelli, Nakai, Frank, and Nishimoto]{denk2023brain2musicreconstructingmusichuman}
Denk, T.~I., Takagi, Y., Matsuyama, T., Agostinelli, A., Nakai, T., Frank, C., and Nishimoto, S.
\newblock Brain2music: Reconstructing music from human brain activity, 2023.
\newblock URL \url{https://arxiv.org/abs/2307.11078}.

\bibitem[Devlin(2018)]{devlin2018bert}
Devlin, J.
\newblock Bert: Pre-training of deep bidirectional transformers for language understanding.
\newblock \emph{arXiv preprint arXiv:1810.04805}, 2018.

\bibitem[Du et~al.(2022)Du, Cheng, Duan, and Ning]{du2022fmri}
Du, B., Cheng, X., Duan, Y., and Ning, H.
\newblock {fMRI} brain decoding and its applications in brain--computer interface: A survey.
\newblock \emph{Brain Sciences}, 12\penalty0 (2):\penalty0 228, 2022.

\bibitem[Epstein et~al.(1999)Epstein, Harris, Stanley, and Kanwisher]{epstein1999parahippocampal}
Epstein, R., Harris, A., Stanley, D., and Kanwisher, N.
\newblock The parahippocampal place area: recognition, navigation, or encoding?
\newblock \emph{Neuron}, 23\penalty0 (1):\penalty0 115--125, 1999.

\bibitem[Epstein \& Ward(2010)Epstein and Ward]{epstein2010reliable}
Epstein, R.~A. and Ward, E.~J.
\newblock How reliable are visual context effects in the parahippocampal place area?
\newblock \emph{Cerebral Cortex}, 20\penalty0 (2):\penalty0 294--303, 2010.

\bibitem[Ferrante et~al.(2023)Ferrante, Ozcelik, Boccato, VanRullen, and Toschi]{ferrante2023brain}
Ferrante, M., Ozcelik, F., Boccato, T., VanRullen, R., and Toschi, N.
\newblock Brain captioning: Decoding human brain activity into images and text.
\newblock \emph{arXiv preprint arXiv:2305.11560}, 2023.

\bibitem[Friederici(2017)]{friederici2017language}
Friederici, A.~D.
\newblock \emph{Language in our brain: The origins of a uniquely human capacity}.
\newblock MIT Press, 2017.

\bibitem[Ganin \& Lempitsky(2015)Ganin and Lempitsky]{ganin2015unsupervised}
Ganin, Y. and Lempitsky, V.
\newblock Unsupervised domain adaptation by backpropagation.
\newblock In \emph{International conference on machine learning}, pp.\  1180--1189. PMLR, 2015.

\bibitem[Ganin et~al.(2016)Ganin, Ustinova, Ajakan, Germain, Larochelle, Laviolette, March, and Lempitsky]{ganin2016domain}
Ganin, Y., Ustinova, E., Ajakan, H., Germain, P., Larochelle, H., Laviolette, F., March, M., and Lempitsky, V.
\newblock Domain-adversarial training of neural networks.
\newblock \emph{Journal of machine learning research}, 17\penalty0 (59):\penalty0 1--35, 2016.

\bibitem[Glasser et~al.(2016)Glasser, Coalson, Robinson, Hacker, Harwell, Yacoub, Ugurbil, Andersson, Beckmann, Jenkinson, et~al.]{glasser2016multi}
Glasser, M.~F., Coalson, T.~S., Robinson, E.~C., Hacker, C.~D., Harwell, J., Yacoub, E., Ugurbil, K., Andersson, J., Beckmann, C.~F., Jenkinson, M., et~al.
\newblock A multi-modal parcellation of human cerebral cortex.
\newblock \emph{Nature}, 536\penalty0 (7615):\penalty0 171--178, 2016.

\bibitem[Golarai et~al.(2007)Golarai, Ghahremani, Whitfield-Gabrieli, Reiss, Eberhardt, Gabrieli, and Grill-Spector]{golarai2007differential}
Golarai, G., Ghahremani, D.~G., Whitfield-Gabrieli, S., Reiss, A., Eberhardt, J.~L., Gabrieli, J.~D., and Grill-Spector, K.
\newblock Differential development of high-level visual cortex correlates with category-specific recognition memory.
\newblock \emph{Nature neuroscience}, 10\penalty0 (4):\penalty0 512--522, 2007.

\bibitem[Gong et~al.(2012)Gong, Shi, Sha, and Grauman]{gong2012geodesic}
Gong, B., Shi, Y., Sha, F., and Grauman, K.
\newblock Geodesic flow kernel for unsupervised domain adaptation.
\newblock In \emph{2012 IEEE conference on computer vision and pattern recognition}, pp.\  2066--2073. IEEE, 2012.

\bibitem[Goyal et~al.(2017)Goyal, Khot, Summers-Stay, Batra, and Parikh]{goyal2017making}
Goyal, Y., Khot, T., Summers-Stay, D., Batra, D., and Parikh, D.
\newblock Making the v in vqa matter: Elevating the role of image understanding in visual question answering.
\newblock In \emph{Proceedings of the IEEE conference on computer vision and pattern recognition}, pp.\  6904--6913, 2017.

\bibitem[Han et~al.(2024{\natexlab{a}})Han, Gong, Zhang, Wang, Zhang, Lin, Qiao, Gao, and Yue]{han2024onellm}
Han, J., Gong, K., Zhang, Y., Wang, J., Zhang, K., Lin, D., Qiao, Y., Gao, P., and Yue, X.
\newblock Onellm: One framework to align all modalities with language.
\newblock In \emph{Proceedings of the IEEE/CVF Conference on Computer Vision and Pattern Recognition}, pp.\  26584--26595, 2024{\natexlab{a}}.

\bibitem[Han et~al.(2024{\natexlab{b}})Han, Wang, Zhai, You, and Yang]{han2024coco}
Han, X., Wang, Y., Zhai, B., You, Q., and Yang, H.
\newblock Coco is" all''you need for visual instruction fine-tuning.
\newblock \emph{arXiv preprint arXiv:2401.08968}, 2024{\natexlab{b}}.

\bibitem[He et~al.(2022)He, Chen, Xie, Li, Doll{\'a}r, and Girshick]{he2022masked}
He, K., Chen, X., Xie, S., Li, Y., Doll{\'a}r, P., and Girshick, R.
\newblock Masked autoencoders are scalable vision learners.
\newblock In \emph{Proceedings of the IEEE/CVF conference on computer vision and pattern recognition}, pp.\  16000--16009, 2022.

\bibitem[Hmamouche et~al.(2024)Hmamouche, Chihab, Kdouri, and Seghrouchni]{hmamouche2024multimodal}
Hmamouche, Y., Chihab, I., Kdouri, L., and Seghrouchni, A. E.~F.
\newblock A multimodal llm for the non-invasive decoding of spoken text from brain recordings.
\newblock \emph{arXiv preprint arXiv:2409.19710}, 2024.

\bibitem[Huang(2024)]{huang2024brainchat}
Huang, W.
\newblock Brainchat: Decoding semantic information from {fMRI} using vision-language pretrained models.
\newblock \emph{arXiv preprint arXiv:2406.07584}, 2024.

\bibitem[Kafle \& Kanan(2017)Kafle and Kanan]{kafle2017analysis}
Kafle, K. and Kanan, C.
\newblock An analysis of visual question answering algorithms.
\newblock In \emph{ICCV}, 2017.

\bibitem[Kan et~al.(2022)Kan, Dai, Cui, Zhang, Guo, and Yang]{kan2022brain}
Kan, X., Dai, W., Cui, H., Zhang, Z., Guo, Y., and Yang, C.
\newblock Brain network transformer.
\newblock \emph{Advances in Neural Information Processing Systems}, 35:\penalty0 25586--25599, 2022.

\bibitem[K{\"o}hler et~al.(2002)K{\"o}hler, Crane, and Milner]{kohler2002differential}
K{\"o}hler, S., Crane, J., and Milner, B.
\newblock Differential contributions of the parahippocampal place area and the anterior hippocampus to human memory for scenes.
\newblock \emph{Hippocampus}, 12\penalty0 (6):\penalty0 718--723, 2002.

\bibitem[Kondratyuk et~al.(2023)Kondratyuk, Yu, Gu, Lezama, Huang, Schindler, Hornung, Birodkar, Yan, Chiu, et~al.]{kondratyuk2023videopoet}
Kondratyuk, D., Yu, L., Gu, X., Lezama, J., Huang, J., Schindler, G., Hornung, R., Birodkar, V., Yan, J., Chiu, M.-C., et~al.
\newblock Videopoet: A large language model for zero-shot video generation.
\newblock \emph{arXiv preprint arXiv:2312.14125}, 2023.

\bibitem[Krause et~al.(2017)Krause, Johnson, Krishna, and Fei-Fei]{krause2017hierarchical}
Krause, J., Johnson, J., Krishna, R., and Fei-Fei, L.
\newblock A hierarchical approach for generating descriptive image paragraphs.
\newblock In \emph{Proceedings of the IEEE conference on computer vision and pattern recognition}, pp.\  317--325, 2017.

\bibitem[Krishna et~al.(2017)Krishna, Zhu, Groth, Johnson, Hata, Kravitz, Chen, Kalantidis, Li, Shamma, et~al.]{krishna2017visual}
Krishna, R., Zhu, Y., Groth, O., Johnson, J., Hata, K., Kravitz, J., Chen, S., Kalantidis, Y., Li, L.-J., Shamma, D.~A., et~al.
\newblock Visual genome: Connecting language and vision using crowdsourced dense image annotations.
\newblock \emph{International journal of computer vision}, 123:\penalty0 32--73, 2017.

\bibitem[Lacan(1988)]{lacan1988seminar}
Lacan, J.
\newblock \emph{The Seminar of Jacques Lacan}.
\newblock WW Norton \& Company, 1988.

\bibitem[Lacan(2001)]{lacan2001ecrits}
Lacan, J.
\newblock \emph{Ecrits: A selection}.
\newblock Routledge, 2001.

\bibitem[Li et~al.(2023)Li, Li, Savarese, and Hoi]{li2023blip}
Li, J., Li, D., Savarese, S., and Hoi, S.
\newblock Blip-2: Bootstrapping language-image pre-training with frozen image encoders and large language models.
\newblock In \emph{International conference on machine learning}, pp.\  19730--19742. PMLR, 2023.

\bibitem[Li et~al.(2018)Li, Tao, Joty, Cai, and Luo]{li2018vqa}
Li, Q., Tao, Q., Joty, S., Cai, J., and Luo, J.
\newblock Vqa-e: Explaining, elaborating, and enhancing your answers for visual questions.
\newblock In \emph{Proceedings of the European Conference on Computer Vision (ECCV)}, pp.\  552--567, 2018.

\bibitem[Li et~al.(2021)Li, Zhou, Dvornek, Zhang, Gao, Zhuang, Scheinost, Staib, Ventola, and Duncan]{li2021braingnn}
Li, X., Zhou, Y., Dvornek, N., Zhang, M., Gao, S., Zhuang, J., Scheinost, D., Staib, L.~H., Ventola, P., and Duncan, J.~S.
\newblock Braingnn: Interpretable brain graph neural network for {fMRI} analysis.
\newblock \emph{Medical Image Analysis}, 74:\penalty0 102233, 2021.

\bibitem[Lin(2004)]{lin2004rouge}
Lin, C.-Y.
\newblock Rouge: A package for automatic evaluation of summaries.
\newblock In \emph{Text summarization branches out}, pp.\  74--81, 2004.

\bibitem[Lin et~al.(2014)Lin, Maire, Belongie, Hays, Perona, Ramanan, Doll{\'a}r, and Zitnick]{lin2014microsoft}
Lin, T.-Y., Maire, M., Belongie, S., Hays, J., Perona, P., Ramanan, D., Doll{\'a}r, P., and Zitnick, C.~L.
\newblock Microsoft coco: Common objects in context.
\newblock In \emph{Computer Vision--ECCV 2014: 13th European Conference, Zurich, Switzerland, September 6-12, 2014, Proceedings, Part V 13}, pp.\  740--755. Springer, 2014.

\bibitem[Liu et~al.(2024)Liu, Jing, Jiang, Wen, Zhu, Li, Pan, Cai, Liu, Zhou, et~al.]{liu2024relationships}
Liu, C., Jing, J., Jiang, J., Wen, W., Zhu, W., Li, Z., Pan, Y., Cai, X., Liu, H., Zhou, Y., et~al.
\newblock Relationships between brain structure-function coupling in normal aging and cognition: A cross-ethnicity population-based study.
\newblock \emph{NeuroImage}, 299:\penalty0 120847, 2024.

\bibitem[Liu et~al.(2023)Liu, Emerson, and Collier]{liu2023visual}
Liu, F., Emerson, G., and Collier, N.
\newblock Visual spatial reasoning.
\newblock \emph{Transactions of the Association for Computational Linguistics}, 11:\penalty0 635--651, 2023.

\bibitem[Luo et~al.(2023)Luo, Henderson, Tarr, and Wehbe]{luo2023brainscuba}
Luo, A.~F., Henderson, M.~M., Tarr, M.~J., and Wehbe, L.
\newblock Brainscuba: Fine-grained natural language captions of visual cortex selectivity.
\newblock \emph{arXiv preprint arXiv:2310.04420}, 2023.

\bibitem[Mai \& Zhang(2023)Mai and Zhang]{mai2023unibrain}
Mai, W. and Zhang, Z.
\newblock Unibrain: Unify image reconstruction and captioning all in one diffusion model from human brain activity.
\newblock \emph{arXiv preprint arXiv:2308.07428}, 2023.

\bibitem[Marino et~al.(2019)Marino, Rastegari, Farhadi, and Mottaghi]{marino2019ok}
Marino, K., Rastegari, M., Farhadi, A., and Mottaghi, R.
\newblock Ok-vqa: A visual question answering benchmark requiring external knowledge.
\newblock In \emph{Proceedings of the IEEE/cvf conference on computer vision and pattern recognition}, pp.\  3195--3204, 2019.

\bibitem[McManus(2002)]{mcmanus2002right}
McManus, I.~C.
\newblock \emph{Right hand, left hand: The origins of asymmetry in brains, bodies, atoms, and cultures}.
\newblock Harvard University Press, 2002.

\bibitem[Miller \& Lacan(2018)Miller and Lacan]{miller2018four}
Miller, J.~A. and Lacan, J.
\newblock \emph{The four fundamental concepts of psycho-analysis}.
\newblock Routledge, 2018.

\bibitem[Murahari et~al.(2019)Murahari, Chattopadhyay, Batra, Parikh, and Das]{murahari2019improving}
Murahari, V., Chattopadhyay, P., Batra, D., Parikh, D., and Das, A.
\newblock Improving generative visual dialog by answering diverse questions.
\newblock \emph{arXiv preprint arXiv:1909.10470}, 2019.

\bibitem[Papineni et~al.(2002)Papineni, Roukos, Ward, and Zhu]{papineni2002bleu}
Papineni, K., Roukos, S., Ward, T., and Zhu, W.-J.
\newblock Bleu: a method for automatic evaluation of machine translation.
\newblock In \emph{Proceedings of the 40th annual meeting of the Association for Computational Linguistics}, pp.\  311--318, 2002.

\bibitem[Qi et~al.(2025)Qi, Dong, Zhang, Geng, Han, Ge, Yi, and Ma]{qi2025shapellm}
Qi, Z., Dong, R., Zhang, S., Geng, H., Han, C., Ge, Z., Yi, L., and Ma, K.
\newblock Shapellm: Universal 3d object understanding for embodied interaction.
\newblock In \emph{European Conference on Computer Vision}, pp.\  214--238. Springer, 2025.

\bibitem[Qiu et~al.(2023)Qiu, Chu, Wang, Zuo, Li, Zhao, and Ying]{qiu2023learning}
Qiu, W., Chu, H., Wang, S., Zuo, H., Li, X., Zhao, Y., and Ying, R.
\newblock Learning high-order relationships of brain regions.
\newblock In \emph{Forty-first International Conference on Machine Learning}, 2023.

\bibitem[Radford et~al.(2019)Radford, Wu, Child, Luan, Amodei, Sutskever, et~al.]{radford2019language}
Radford, A., Wu, J., Child, R., Luan, D., Amodei, D., Sutskever, I., et~al.
\newblock Language models are unsupervised multitask learners.
\newblock \emph{OpenAI blog}, 1\penalty0 (8):\penalty0 9, 2019.

\bibitem[Radford et~al.(2021)Radford, Kim, Hallacy, Ramesh, Goh, Agarwal, Sastry, Askell, Mishkin, Clark, et~al.]{radford2021learning}
Radford, A., Kim, J.~W., Hallacy, C., Ramesh, A., Goh, G., Agarwal, S., Sastry, G., Askell, A., Mishkin, P., Clark, J., et~al.
\newblock Learning transferable visual models from natural language supervision.
\newblock In \emph{International conference on machine learning}, pp.\  8748--8763. PMLR, 2021.

\bibitem[Ranganath et~al.(2004)Ranganath, DeGutis, and D'Esposito]{ranganath2004category}
Ranganath, C., DeGutis, J., and D'Esposito, M.
\newblock Category-specific modulation of inferior temporal activity during working memory encoding and maintenance.
\newblock \emph{Cognitive Brain Research}, 20\penalty0 (1):\penalty0 37--45, 2004.

\bibitem[Ren et~al.(2015)Ren, Kiros, and Zemel]{ren2015exploring}
Ren, M., Kiros, R., and Zemel, R.
\newblock Exploring models and data for image question answering.
\newblock \emph{Advances in neural information processing systems}, 28, 2015.

\bibitem[Robertson(2002)]{robertson2002memory}
Robertson, L.~T.
\newblock Memory and the brain.
\newblock \emph{Journal of dental education}, 66\penalty0 (1):\penalty0 30--42, 2002.

\bibitem[Rolls et~al.(2020)Rolls, Huang, Lin, Feng, and Joliot]{rolls2020automated}
Rolls, E.~T., Huang, C.-C., Lin, C.-P., Feng, J., and Joliot, M.
\newblock Automated anatomical labelling atlas 3.
\newblock \emph{Neuroimage}, 206:\penalty0 116189, 2020.

\bibitem[Schultz et~al.(2003)Schultz, Grelotti, Klin, Kleinman, Van~der Gaag, Marois, and Skudlarski]{schultz2003role}
Schultz, R.~T., Grelotti, D.~J., Klin, A., Kleinman, J., Van~der Gaag, C., Marois, R., and Skudlarski, P.
\newblock The role of the fusiform face area in social cognition: implications for the pathobiology of autism.
\newblock \emph{Philosophical Transactions of the Royal Society of London. Series B: Biological Sciences}, 358\penalty0 (1430):\penalty0 415--427, 2003.

\bibitem[Schwenk et~al.(2022)Schwenk, Khandelwal, Clark, Marino, and Mottaghi]{schwenk2022okvqa}
Schwenk, D., Khandelwal, A., Clark, C., Marino, K., and Mottaghi, R.
\newblock A-okvqa: A benchmark for visual question answering using world knowledge.
\newblock In \emph{European conference on computer vision}, pp.\  146--162. Springer, 2022.

\bibitem[Scotti et~al.(2024{\natexlab{a}})Scotti, Banerjee, Goode, Shabalin, Nguyen, Dempster, Verlinde, Yundler, Weisberg, Norman, et~al.]{scotti2024reconstructing}
Scotti, P., Banerjee, A., Goode, J., Shabalin, S., Nguyen, A., Dempster, A., Verlinde, N., Yundler, E., Weisberg, D., Norman, K., et~al.
\newblock Reconstructing the mind's eye: {fMRI}-to-image with contrastive learning and diffusion priors.
\newblock \emph{Advances in Neural Information Processing Systems}, 36, 2024{\natexlab{a}}.

\bibitem[Scotti et~al.(2024{\natexlab{b}})Scotti, Tripathy, Villanueva, Kneeland, Chen, Narang, Santhirasegaran, Xu, Naselaris, Norman, et~al.]{scotti2024mindeye2}
Scotti, P.~S., Tripathy, M., Villanueva, C. K.~T., Kneeland, R., Chen, T., Narang, A., Santhirasegaran, C., Xu, J., Naselaris, T., Norman, K.~A., et~al.
\newblock Mindeye2: Shared-subject models enable {fMRI}-to-image with 1 hour of data.
\newblock \emph{arXiv preprint arXiv:2403.11207}, 2024{\natexlab{b}}.

\bibitem[Shin et~al.(2016)Shin, Ushiku, and Harada]{shin2016color}
Shin, A., Ushiku, Y., and Harada, T.
\newblock The color of the cat is gray: 1 million full-sentences visual question answering (fsvqa).
\newblock \emph{arXiv preprint arXiv:1609.06657}, 2016.

\bibitem[Stenning \& Van~Lambalgen(2012)Stenning and Van~Lambalgen]{stenning2012human}
Stenning, K. and Van~Lambalgen, M.
\newblock \emph{Human reasoning and cognitive science}.
\newblock MIT Press, 2012.

\bibitem[Sun et~al.(2024)Sun, Li, Chen, and Moens]{sun2024neurocinedecodingvividvideo}
Sun, J., Li, M., Chen, Z., and Moens, M.-F.
\newblock Neurocine: Decoding vivid video sequences from human brain activties, 2024.
\newblock URL \url{https://arxiv.org/abs/2402.01590}.

\bibitem[Takagi \& Nishimoto(2023)Takagi and Nishimoto]{takagi2023high}
Takagi, Y. and Nishimoto, S.
\newblock High-resolution image reconstruction with latent diffusion models from human brain activity.
\newblock In \emph{Proceedings of the IEEE/CVF Conference on Computer Vision and Pattern Recognition}, pp.\  14453--14463, 2023.

\bibitem[Tancik et~al.(2020)Tancik, Srinivasan, Mildenhall, Fridovich-Keil, Raghavan, Singhal, Ramamoorthi, Barron, and Ng]{tancik2020fourier}
Tancik, M., Srinivasan, P., Mildenhall, B., Fridovich-Keil, S., Raghavan, N., Singhal, U., Ramamoorthi, R., Barron, J., and Ng, R.
\newblock Fourier features let networks learn high frequency functions in low dimensional domains.
\newblock \emph{Advances in neural information processing systems}, 33:\penalty0 7537--7547, 2020.

\bibitem[Tolstikhin et~al.(2021)Tolstikhin, Houlsby, Kolesnikov, Beyer, Zhai, Unterthiner, Yung, Steiner, Keysers, Uszkoreit, et~al.]{tolstikhin2021mlp}
Tolstikhin, I.~O., Houlsby, N., Kolesnikov, A., Beyer, L., Zhai, X., Unterthiner, T., Yung, J., Steiner, A., Keysers, D., Uszkoreit, J., et~al.
\newblock Mlp-mixer: An all-mlp architecture for vision.
\newblock \emph{Advances in neural information processing systems}, 34:\penalty0 24261--24272, 2021.

\bibitem[Toneva \& Wehbe(2019)Toneva and Wehbe]{toneva2019interpreting}
Toneva, M. and Wehbe, L.
\newblock Interpreting and improving natural-language processing (in machines) with natural language-processing (in the brain).
\newblock \emph{Advances in neural information processing systems}, 32, 2019.

\bibitem[Tsantani et~al.(2021)Tsantani, Kriegeskorte, Storrs, Williams, McGettigan, and Garrido]{tsantani2021ffa}
Tsantani, M., Kriegeskorte, N., Storrs, K., Williams, A.~L., McGettigan, C., and Garrido, L.
\newblock Ffa and ofa encode distinct types of face identity information.
\newblock \emph{Journal of Neuroscience}, 41\penalty0 (9):\penalty0 1952--1969, 2021.

\bibitem[Tunik et~al.(2007)Tunik, Rice, Hamilton, and Grafton]{tunik2007beyond}
Tunik, E., Rice, N.~J., Hamilton, A., and Grafton, S.~T.
\newblock Beyond grasping: representation of action in human anterior intraparietal sulcus.
\newblock \emph{Neuroimage}, 36:\penalty0 T77--T86, 2007.

\bibitem[Vaswani(2017)]{vaswani2017attention}
Vaswani, A.
\newblock Attention is all you need.
\newblock \emph{Advances in Neural Information Processing Systems}, 2017.

\bibitem[Vedantam et~al.(2015)Vedantam, Lawrence~Zitnick, and Parikh]{vedantam2015cider}
Vedantam, R., Lawrence~Zitnick, C., and Parikh, D.
\newblock Cider: Consensus-based image description evaluation.
\newblock In \emph{Proceedings of the IEEE conference on computer vision and pattern recognition}, pp.\  4566--4575, 2015.

\bibitem[Wade \& Swanston(2013)Wade and Swanston]{wade2013visual}
Wade, N. and Swanston, M.
\newblock \emph{Visual perception: An introduction}.
\newblock Psychology Press, 2013.

\bibitem[Wang et~al.(2022)Wang, Yang, Hu, Li, Lin, Gan, Liu, Liu, and Wang]{wang2022git}
Wang, J., Yang, Z., Hu, X., Li, L., Lin, K., Gan, Z., Liu, Z., Liu, C., and Wang, L.
\newblock Git: A generative image-to-text transformer for vision and language.
\newblock \emph{arXiv preprint arXiv:2205.14100}, 2022.

\bibitem[Wang et~al.(2023{\natexlab{a}})Wang, Meng, Weng, He, Wu, and Jiang]{wang2023see}
Wang, J., Meng, L., Weng, Z., He, B., Wu, Z., and Jiang, Y.-G.
\newblock To see is to believe: Prompting gpt-4v for better visual instruction tuning.
\newblock \emph{arXiv preprint arXiv:2311.07574}, 2023{\natexlab{a}}.

\bibitem[Wang et~al.(2024{\natexlab{a}})Wang, Liu, Tan, and Wang]{wang2024mindbridge}
Wang, S., Liu, S., Tan, Z., and Wang, X.
\newblock Mindbridge: A cross-subject brain decoding framework.
\newblock In \emph{Proceedings of the IEEE/CVF Conference on Computer Vision and Pattern Recognition}, pp.\  11333--11342, 2024{\natexlab{a}}.

\bibitem[Wang et~al.(2023{\natexlab{b}})Wang, Lv, Yu, Hong, Qi, Wang, Ji, Yang, Zhao, Song, et~al.]{wang2023cogvlm}
Wang, W., Lv, Q., Yu, W., Hong, W., Qi, J., Wang, Y., Ji, J., Yang, Z., Zhao, L., Song, X., et~al.
\newblock Cogvlm: Visual expert for pretrained language models.
\newblock \emph{arXiv preprint arXiv:2311.03079}, 2023{\natexlab{b}}.

\bibitem[Wang et~al.(2024{\natexlab{b}})Wang, Zhao, Zhou, and Nachev]{wang2024unibrain}
Wang, Z., Zhao, Z., Zhou, L., and Nachev, P.
\newblock Unibrain: A unified model for cross-subject brain decoding.
\newblock \emph{arXiv preprint arXiv:2412.19487}, 2024{\natexlab{b}}.

\bibitem[Xia et~al.(2024)Xia, de~Charette, {\"O}ztireli, and Xue]{xia2024umbrae}
Xia, W., de~Charette, R., {\"O}ztireli, C., and Xue, J.-H.
\newblock Umbrae: Unified multimodal decoding of brain signals.
\newblock \emph{arXiv preprint arXiv:2404.07202}, 2024.

\bibitem[Xiao et~al.(2023)Xiao, Wang, Jin, Feng, Chen, Huang, and Zhao]{xiao2023spa}
Xiao, Z., Wang, H., Jin, Y., Feng, L., Chen, G., Huang, F., and Zhao, J.
\newblock Spa: A graph spectral alignment perspective for domain adaptation.
\newblock \emph{Advances in Neural Information Processing Systems}, 36:\penalty0 37252--37272, 2023.

\bibitem[Xu et~al.(2023)Xu, Wang, Wang, Chen, Pang, and Lin]{xu2023pointllm}
Xu, R., Wang, X., Wang, T., Chen, Y., Pang, J., and Lin, D.
\newblock Pointllm: Empowering large language models to understand point clouds.
\newblock \emph{arXiv preprint arXiv:2308.16911}, 2023.

\bibitem[Xu(2005)]{xu2005revisiting}
Xu, Y.
\newblock Revisiting the role of the fusiform face area in visual expertise.
\newblock \emph{Cerebral Cortex}, 15\penalty0 (8):\penalty0 1234--1242, 2005.

\bibitem[Yu et~al.(2022)Yu, Wang, Vasudevan, Yeung, Seyedhosseini, and Wu]{yu2022coca}
Yu, J., Wang, Z., Vasudevan, V., Yeung, L., Seyedhosseini, M., and Wu, Y.
\newblock Coca: Contrastive captioners are image-text foundation models.
\newblock \emph{arXiv preprint arXiv:2205.01917}, 2022.

\bibitem[Zhang et~al.(2023{\natexlab{a}})Zhang, Li, and Bing]{zhang2023video}
Zhang, H., Li, X., and Bing, L.
\newblock Video-llama: An instruction-tuned audio-visual language model for video understanding.
\newblock \emph{arXiv preprint arXiv:2306.02858}, 2023{\natexlab{a}}.

\bibitem[Zhang et~al.(2023{\natexlab{b}})Zhang, Dong, Wang, Cao, Xu, Ouyang, Zhao, Duan, Zhang, Ding, et~al.]{zhang2023internlm}
Zhang, P., Dong, X., Wang, B., Cao, Y., Xu, C., Ouyang, L., Zhao, Z., Duan, H., Zhang, S., Ding, S., et~al.
\newblock Internlm-xcomposer: A vision-language large model for advanced text-image comprehension and composition.
\newblock \emph{arXiv preprint arXiv:2309.15112}, 2023{\natexlab{b}}.

\bibitem[Zhang et~al.(2023{\natexlab{c}})Zhang, Liang, Wang, Wang, Gao, Sui, Xin, Feng, Guo, and Wen]{zhang2023abnormal}
Zhang, X., Liang, C., Wang, N., Wang, Y., Gao, Y., Sui, C., Xin, H., Feng, M., Guo, L., and Wen, H.
\newblock Abnormal whole-brain voxelwise structure-function coupling and its association with cognitive dysfunction in patients with different cerebral small vessel disease burdens.
\newblock \emph{Frontiers in Aging Neuroscience}, 15:\penalty0 1148738, 2023{\natexlab{c}}.

\bibitem[Zheng et~al.(2023)Zheng, Chiang, Sheng, Zhuang, Wu, Zhuang, Lin, Li, Li, Xing, et~al.]{zheng2023judging}
Zheng, L., Chiang, W.-L., Sheng, Y., Zhuang, S., Wu, Z., Zhuang, Y., Lin, Z., Li, Z., Li, D., Xing, E., et~al.
\newblock Judging llm-as-a-judge with mt-bench and chatbot arena.
\newblock \emph{Advances in Neural Information Processing Systems}, 36:\penalty0 46595--46623, 2023.

\bibitem[Zhu et~al.(2020)Zhu, Su, Lu, Li, Wang, and Dai]{zhu2020deformable}
Zhu, X., Su, W., Lu, L., Li, B., Wang, X., and Dai, J.
\newblock Deformable detr: Deformable transformers for end-to-end object detection.
\newblock \emph{arXiv preprint arXiv:2010.04159}, 2020.

\end{thebibliography}
\bibliographystyle{icml2025}


\appendix
\onecolumn
\section{Dataset Details}
\label{app:dataset}

\subsection{Details of each dataset in brain instruction tuning}
In this section, we give a brief description of each source of our brain instruction datasets as well as examples from them.

\setlength{\LTcapwidth}{\textwidth}

{\scriptsize 
\begin{longtable}{>{\raggedright\arraybackslash}m{0.2\textwidth}m{0.2\textwidth}m{0.5\textwidth}}
\caption{Dataset details and examples.}
\label{tab:dataset_details}\\

\toprule
\textbf{Dataset} & \textbf{Description} & \textbf{Example}\\
\midrule
\endfirsthead

\toprule
\textbf{Dataset} & \textbf{Description} & \textbf{Example}\\
\midrule
\endhead

\midrule
\multicolumn{3}{r}{Continued on next page}\\
\endfoot

\bottomrule
\endlastfoot

\begin{minipage}[c]{\linewidth}
    \vspace{0.5em}
    Previous Caption
    \vspace{0.5em}
\end{minipage} & 
\begin{minipage}[c]{\linewidth}
    \vspace{0.5em}
    Generating a one-sentence caption of the image that the subject previously saw.
    \vspace{0.5em}
\end{minipage} &
\begin{minipage}[c]{\linewidth}
    \centering
    \begin{minipage}[c]{0.3\linewidth}
        \includegraphics[width=0.7\linewidth]{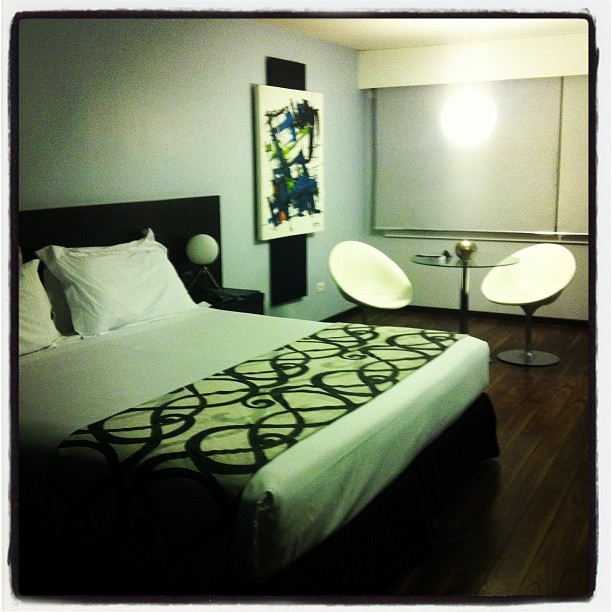}
    \end{minipage}%
    \hfill
    \begin{minipage}[c]{0.65\linewidth}
        A neat bedroom pairs modern chairs with a glass table.
    \end{minipage}
\end{minipage}\\
\midrule

\begin{minipage}[c]{\linewidth}
    \vspace{0.5em}
    \makecell[l]{COCO Caption\\\cite{chen2015microsoft}}
    \vspace{0.5em}
\end{minipage} & 
\begin{minipage}[c]{\linewidth}
    \vspace{0.5em}
    Generate a one-sentence caption of the image the subject currently sees.
    \vspace{0.5em}
\end{minipage} &
\begin{minipage}[c]{\linewidth}
    \centering
    \begin{minipage}[c]{0.3\linewidth}
        \includegraphics[width=0.7\linewidth]{figures/table5_example/coco-caption.jpg}
    \end{minipage}%
    \hfill
    \begin{minipage}[c]{0.65\linewidth}
        The pedestrian is walking down the side of the highway by the bus.
    \end{minipage}
\end{minipage}\\
\midrule

\begin{minipage}[c]{\linewidth}
    \vspace{0.5em}
Image Paragraph Captioning\\ \cite{krause2017hierarchical}
\vspace{0.5em}
\end{minipage} & 
\begin{minipage}[c]{\linewidth}
    \vspace{0.5em}
 Generate a one-paragraph caption of the image the subject currently sees    \vspace{0.5em}
\end{minipage} &
\begin{minipage}[c]{\linewidth}
    \centering
    \begin{minipage}[c]{0.3\linewidth}
        \includegraphics[width=0.7\linewidth]{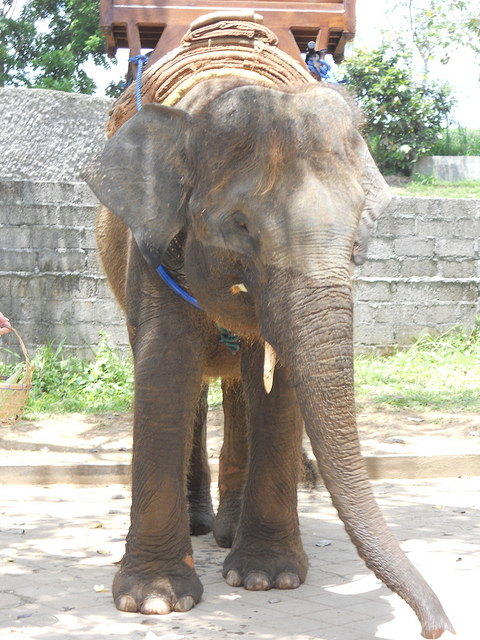}
    \end{minipage}%
    \hfill
    \begin{minipage}[c]{0.65\linewidth}
        An elephant with a harness, and a seat on his back is seen in a dirt field on a sunny day. The seat on the back of the elephant is brown and is tied with ropes. The sun is shining through the trees onto the ground below. Behind the elephant there is a cinder block wall with grass growing in front of the wall. There are trees behind the wall.    \end{minipage}
\end{minipage}\\
\midrule

\begin{minipage}[c]{\linewidth}
    \vspace{0.5em}
COCO QA\\\cite{ren2015exploring}
\vspace{0.5em}
\end{minipage} & 
\begin{minipage}[c]{\linewidth}
    \vspace{0.5em}
Answer questions according to the image.
\vspace{0.5em}
\end{minipage} &
\begin{minipage}[c]{\linewidth}
    \centering
    \begin{minipage}[c]{0.3\linewidth}
        \includegraphics[width=0.7\linewidth]{figures/table5_example/coco-qa.jpg}
    \end{minipage}%
    \hfill
    \begin{minipage}[c]{0.65\linewidth}
        [Q]: what eats leaves from a basket in an enclosure? [A]: Giraffe.
    \end{minipage}
\end{minipage}\\
\midrule

\begin{minipage}[c]{\linewidth}
    \vspace{0.5em}
Visual Genome QA\\\cite{krishna2017visual}
\vspace{0.5em}
\end{minipage} & 
\begin{minipage}[c]{\linewidth}
    \vspace{0.5em}
Answer image-based questions that require richer semantic understanding of the image than COCO-QA.
\vspace{0.5em}
\end{minipage} &
\begin{minipage}[c]{\linewidth}
    \centering
    \begin{minipage}[c]{0.3\linewidth}
        \includegraphics[width=0.7\linewidth]{figures/table5_example/visual-genome.jpg}
    \end{minipage}%
    \hfill
    \begin{minipage}[c]{0.65\linewidth}
        [Q]: Where was the photo taken? [A]: In an office.
    \end{minipage}
\end{minipage}\\
\midrule

\begin{minipage}[c]{\linewidth}
    \vspace{0.5em}
VQAv2\\\cite{goyal2017making}
\vspace{0.5em}
\end{minipage} & 
\begin{minipage}[c]{\linewidth}
    \vspace{0.5em}
 Answer image-based questions with better equality and diversity than COCO-QA.
 
 \vspace{0.5em}
\end{minipage} &
\begin{minipage}[c]{\linewidth}
    \centering
    \begin{minipage}[c]{0.3\linewidth}
        \includegraphics[width=0.7\linewidth]{figures/table5_example/vqa-v2.jpg}
    \end{minipage}%
    \hfill
    \begin{minipage}[c]{0.65\linewidth}
        [Q]: What are the two white letters? [A]: hu
    \end{minipage}
\end{minipage}\\
\midrule

\begin{minipage}[c]{\linewidth}
    \vspace{0.5em}
OK-VQA\\\cite{marino2019ok}
\vspace{0.5em}
\end{minipage} & 
\begin{minipage}[c]{\linewidth}
    \vspace{0.5em}
 Answer image-based questions that require external knowledge beyond the image itself.
\vspace{0.5em}
\end{minipage} &
\begin{minipage}[c]{\linewidth}
    \centering
    \begin{minipage}[c]{0.3\linewidth}
        \includegraphics[width=0.7\linewidth]{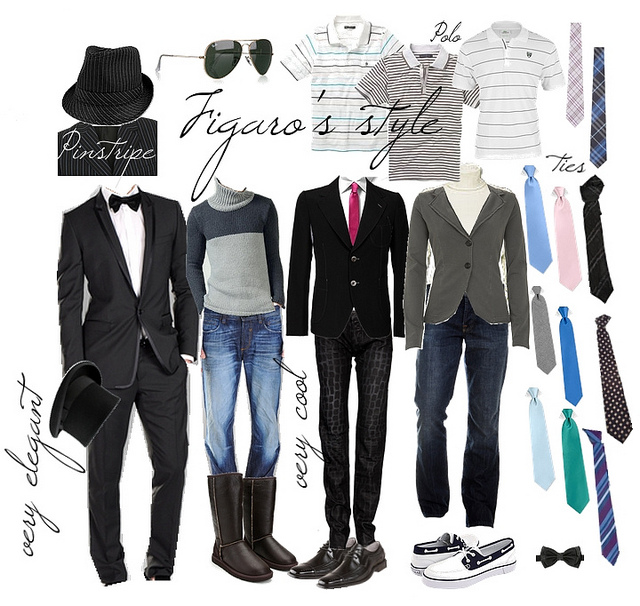}
    \end{minipage}%
    \hfill
    \begin{minipage}[c]{0.65\linewidth}
        [Q]: What part of the body do you wear the rightmost objects on? [A]: Neck.
    \end{minipage}
\end{minipage}\\
\midrule

\begin{minipage}[c]{\linewidth}
    \vspace{0.5em}
ST-VQA\\\cite{biten2019scene}
\vspace{0.5em}
\end{minipage} & 
\begin{minipage}[c]{\linewidth}
    \vspace{0.5em}
 Answer questions of high-level semantic information present in images as the textual cue
 \vspace{0.5em}
\end{minipage} &
\begin{minipage}[c]{\linewidth}
    \centering
    \begin{minipage}[c]{0.3\linewidth}
        \includegraphics[width=0.7\linewidth]{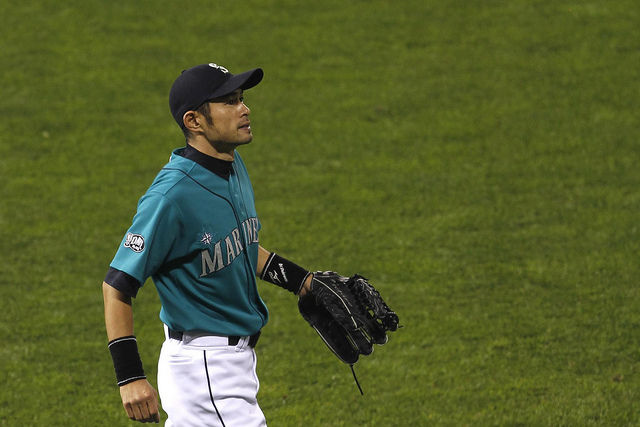}
    \end{minipage}%
    \hfill
    \begin{minipage}[c]{0.65\linewidth}
        [Q]: What is written on the front of the shirt? [A]: Marine.
    \end{minipage}
\end{minipage}\\
\midrule

\begin{minipage}[c]{\linewidth}
    \vspace{0.5em}
TallyQA\\\cite{acharya2019tallyqa}
\vspace{0.5em}
\end{minipage} & 
\begin{minipage}[c]{\linewidth}
    \vspace{0.5em}
Count objects present in images.
\vspace{0.5em}
\end{minipage} &
\begin{minipage}[c]{\linewidth}
    \centering
    \begin{minipage}[c]{0.3\linewidth}
        \includegraphics[width=0.7\linewidth]{figures/table5_example/tally-qa.jpg}
    \end{minipage}%
    \hfill
    \begin{minipage}[c]{0.65\linewidth}
        [Q]: How many bats on the wall? [A]: 2
    \end{minipage}
\end{minipage}\\
\midrule

\begin{minipage}[c]{\linewidth}
    \vspace{0.5em}
VQA-E\\\cite{li2018vqa}
\vspace{0.5em}
\end{minipage} & 
\begin{minipage}[c]{\linewidth}
    \vspace{0.5em}
Answer questions and generate corresponding explanations for an image-based question.
\vspace{0.5em}
\end{minipage} &
\begin{minipage}[c]{\linewidth}
    \centering
    \begin{minipage}[c]{0.3\linewidth}
        \includegraphics[width=0.7\linewidth]{figures/table5_example/vqa-e.jpg}
    \end{minipage}%
    \hfill
    \begin{minipage}[c]{0.65\linewidth}
        [Q]: Are the people going for a walk in the forest? [A]: Yes. Here is the explanation: A picture of the land, trees, and people passing by as they ride in a vehicle.
    \end{minipage}
\end{minipage}\\
\midrule

\begin{minipage}[c]{\linewidth}
    \vspace{0.5em}
A-OKVQA\\\cite{schwenk2022okvqa}
\vspace{0.5em}
\end{minipage} & 
\begin{minipage}[c]{\linewidth}
    \vspace{0.5em}
Answer multiple-choice questions.
\vspace{0.5em}
\end{minipage} &
\begin{minipage}[c]{\linewidth}
    \centering
    \begin{minipage}[c]{0.3\linewidth}
        \includegraphics[width=0.7\linewidth]{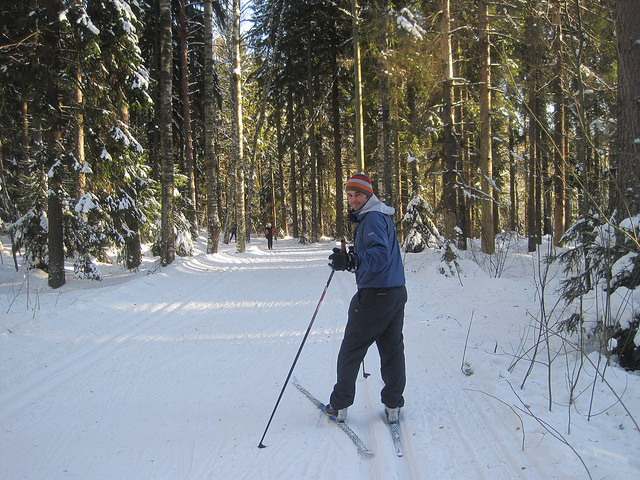}
    \end{minipage}%
    \hfill
    \begin{minipage}[c]{0.65\linewidth}
        [Q]: What season is up next? Multiple Choices: A. autumn B. spring C. summer D. winter [A]: B
    \end{minipage}
\end{minipage}\\
\midrule

\begin{minipage}[c]{\linewidth}
    \vspace{0.5em}
FSVQA\\\cite{shin2016color}
\vspace{0.5em}
\end{minipage} & 
\begin{minipage}[c]{\linewidth}
    \vspace{0.5em}
Answer the questions in full sentences.
\vspace{0.5em}
\end{minipage} &
\begin{minipage}[c]{\linewidth}
    \centering
    \begin{minipage}[c]{0.3\linewidth}
        \includegraphics[width=0.7\linewidth]{figures/table5_example/fsvqa.jpg}
    \end{minipage}%
    \hfill
    \begin{minipage}[c]{0.65\linewidth}
        [Q]: Is this an area that is more populated with pedestrians than cars? [A]: Yes, this is an area that is more populated with pedestrians than cars.
    \end{minipage}
\end{minipage}\\
\midrule

\begin{minipage}[c]{\linewidth}
    \vspace{0.5em}
VisDial\\\cite{murahari2019improving}

\vspace{0.5em}
\end{minipage} & 
\begin{minipage}[c]{\linewidth}
    \vspace{0.5em}
Generate answers for image-based questions in a multi-turn dialogue.     \vspace{0.5em}
\end{minipage} &
\begin{minipage}[c]{\linewidth}
    \centering
    \begin{minipage}[c]{0.3\linewidth}
        \includegraphics[width=0.7\linewidth]{figures/table5_example/visdial.jpg}
    \end{minipage}%
    \hfill
    \begin{minipage}[c]{0.65\linewidth}
        [Q1]: Is the photo in color? [A1]: No. [Q2]: Is the man wearing glasses [A2]: Can't tell. [Q3]: How many horses are there? [A3]: 2 
    \end{minipage}
\end{minipage}\\
\midrule

\begin{minipage}[c]{\linewidth}
    \vspace{0.5em}
LLava Instruction 150K\\\cite{liu2023visual}
\vspace{0.5em}
\end{minipage} & 
\begin{minipage}[c]{\linewidth}
    \vspace{0.5em}
Generate answers for object-level and scene-level answering or reasoning questions for single or multi-round conversations. 
\vspace{0.5em}
\end{minipage} &
\begin{minipage}[c]{\linewidth}
    \centering
    \begin{minipage}[c]{0.3\linewidth}
        \includegraphics[width=0.7\linewidth]{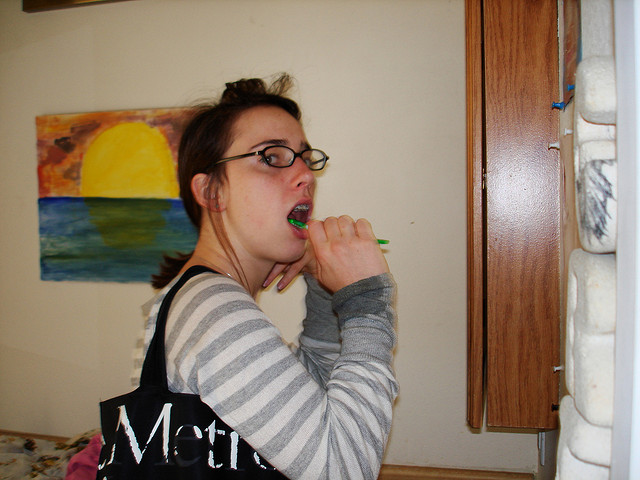}
    \end{minipage}%
    \hfill
    \begin{minipage}[c]{0.65\linewidth}
        [Q1]: What is the main activity of the person in the fMRI? [A1]: The main activity of the person in the fMRI is brushing her teeth. [Q2]: What should be considered when maintaining oral hygiene in a public setting? [A2]: When maintaining oral hygiene in a public setting (as depicted in the fMRI by a woman brushing her teeth with a green toothbrush while carrying a shoulder bag), factors such as cleanliness, privacy, time management, and personal hygiene supplies should be taken into account...
    \end{minipage}
\end{minipage}\\
\midrule

\begin{minipage}[c]{\linewidth}
    \vspace{0.5em}
LVIS Instruct4V\\\cite{wang2023see}

\vspace{0.5em}
\end{minipage} & 
\begin{minipage}[c]{\linewidth}
    \vspace{0.8em}
 Similar to LLava Instruct4V but is finer-grained and of higher quality.
 \vspace{0.5em}
\end{minipage} &
\begin{minipage}[c]{\linewidth}
    \centering
    \begin{minipage}[c]{0.3\linewidth}
        \includegraphics[width=0.7\linewidth]{figures/table5_example/lvis.jpg}
    \end{minipage}%
    \hfill
    \begin{minipage}[c]{0.65\linewidth}
        [Q1]: What is the primary object in the fMRI? [A1]: The primary object in the fMRI is a bird. [Q2]: Can you describe the color and appearance of the bird? [A2]: The bird has a deep brown body with a darker head. Its eyes are distinctively surrounded by a yellowish hue. The bird also possesses a blue-tinted beak. [Q3]: How many birds are in the picture?' [A3]: There is only one bird in the picture...
    \end{minipage}
\end{minipage}\\
\midrule

\begin{minipage}[c]{\linewidth}
    \vspace{0.5em}
TDIUC (only used in downstream)\\\cite{kafle2017analysis}
\vspace{0.5em}
\end{minipage} & 
\begin{minipage}[c]{\linewidth}
    \vspace{0.5em}
Generate answers for image-based questions from 12 different task-directed question types such as object presence and utility/affordance. 
\vspace{0.5em}
\end{minipage} &
\begin{minipage}[c]{\linewidth}
    \centering
    \begin{minipage}[c]{0.3\linewidth}
        \includegraphics[width=0.7\linewidth]{figures/tdiuc.jpg}
    \end{minipage}%
    \hfill
    \begin{minipage}[c]{0.65\linewidth}
        [Q]: What object can be thrown? [A]: Baseball
    \end{minipage}
\end{minipage}\\

\end{longtable}
}  

\subsection{Dataset statistics}
Table~\ref{tab:num_voxels} summarizes the statistics of each subject in the natural scene dataset \cite{allen2022massive}. 
\begin{table}[htbp]
    \caption{Number of voxels for each subject.}
    \label{tab:num_voxels}
    \vskip 0.1in
    \centering
    \begin{tabular}{c|cccccccc}
    \toprule
        subject & 1 & 2 & 3 & 4 & 5 & 6 & 7 & 8 \\
        \midrule
        \#(input voxels) & $15724$ & $14278$ & $15226$ & $13153$ & $13039$ & $17907$ & $126682$ & $14386$ \\
        \#(samples) & $30000$ & $30000$ & $24000$ & $22500$ & $30000$ & $24000$ & $30000$ & $22500$ \\
        \bottomrule
    \end{tabular}
\end{table}



\subsection{Instructions}
Table~\ref{tab:ift} lists instructions for each dataset (i.e. $X_\text{inst}$). Some instructions are inspired by \cite{han2024coco}. For question-answering tasks, the instructions are appended after the question in a new line.

\begin{table}[htbp]
    \caption{Instruction template and statistics of the BIT dataset. "$-$" indicates that the instruction is embedded within the conversation and does not require an additional one. Note that not all conversations are associated with fMRI recordings as only a subset of MSCOCO images were used as stimuli in the study. Consequently, the number of usable conversations in practice will be lower.}
    \label{tab:ift}
    \vskip 0.1in
    \centering
    \resizebox{\linewidth}{!}{
    \begin{tabular}{ccp{30em}}
    \toprule
    Dataset & \#(conversations) &Instruction  \\
    \midrule
    system prompt & / & You are a helpful agent that decodes the brain activity of a person looking at an image. \\
    Previous Caption & $149,875$ &Please describe the image the subject saw previously. \\
     COCO Caption & $616,767$ &Please describe the fMRI as simply as possible.  \\
     Image Paragraph Captioning & $9,598$ &Describe the fMRI in one paragraph.  \\
     COCO QA & $117,684$ & Answer the question with a short phrase. \\
     Visual Genome QA & $676,116$ & Answer the question with a short phrase. \\
     VQAv2 & $6,581,110$ & Answer the question with a short phrase.\\
     OK-VQA & $140,550$ & Answer the question with a short phrase. \\
     ST-VQA & $29,751$ &Answer the question with a short phrase. \\
     TallyQA & $238,056$ &Answer the question with a number.  \\
     VQA-E & $2,697,860$ &Answer with a short phrase and provide explanation for your answer. \\
     A-OKVQA & $18,201$ &Answer with the option's letter from the given choices and provide explanation for your choice. \\
     FSVQA & $369,861$ & Answer the question in a full sentence. \\
     VisDial & $125,351$ & $-$\\ 
     Llava Instruct 150K & $157,712$ & $-$ \\
     LVIS Instruct4V & $222,711$ & $-$\\
     \bottomrule
    \end{tabular}
    }
\end{table}

\section{Implementation details} 
\label{app:impl}
We choose Vicuna-7b \cite{zheng2023judging} as our backbone LLM. During the brain instruction tuning stage, We use AdamW as the optimizer, with the learning rate of $1\times 10^{-3}$, weight decay of $0.01$ and $\beta_1=0.9$, $\beta_2=0.999$. We do not use a learning rate scheduler. We set the batch size to $64$. The instruction tuning is conducted on a machine with $8 \times$ L40S GPUs for $8$ days. And each downstream fine-tuning is conducted on a single L40S GPU with a $1\times 10^{-4}$ learning rate and $48$ batch size. For generations, we have adopted the greedy decoding strategy.

We use $128$ as the number of fMRI tokens. We use a $4$-layer MLP in the fMRI encoder. We set the number of queries to $1024$. The dimension of the query embeddings is $128$.


        

\section{Computational Complexity and Runtime Analysis}
\label{app:complexity}
In the neuroscience-informed attention, the complexity of the dot product between queries and keys is $O(dNN_q)$. The complexity of the aggregation of values is $O(NN_q)$, which is negligible. The MLP maps the hidden representation of dimension $N_q$ to $L \times d$, therefore its complexity is $dLN_q$. Therefore, the complexity of the fMRI encoder is $O(dNN_q + dLN_q) = O(dN_q(L + N)) = O(dN_qN)$ given that $L \ll N$.

Table~\ref{tab:computation} summarizes the computational complexity and runtime comparison of different models. Despite its superior performance, our model introduces only a marginal increase in encoder-side computational cost compared to other baselines. Notably, the majority of inference time and complexity arises from the LLM component, which is shared across all models. This highlights that the design choices in the encoder, while crucial for performance, do not significantly affect runtime.

\begin{table}[htbp]
    \centering
        \caption{Computational complexity and runtime of each encoder and the LLM. For the first 4 lines, we only time the encoder part, excluding the LLM. The runtime of the LLM forward pass is reported separately in the last row.}
    \label{tab:computation}
    \vskip 0.1in
    \begin{tabular}{cccll}
    \toprule
    Module	&Forward Pass (ms/sample)	&Complexity	&Notation Explanation	&Approximate Values\\
    \midrule
UMBRAE	&$0.0109$	&$O(dN + dN_q^2)$& \begin{tabular}{l}$d$: hidden size\\$N$: number of voxels\\$N_q$: number of query tokens\end{tabular}	& \begin{tabular}{l}$d = 100$\\$N = 15000$\\$N_q = 200$\end{tabular}\\
\midrule
MindBridge &$0.0048$ & $O(dN)$ & \begin{tabular}{l}$d$: hidden size\\$N$: number of voxels\end{tabular} & \begin{tabular}{l}$d = 1000$\\$N = 15000$\end{tabular}\\
\midrule
UniBrain &$0.0175$ & $O(dGK+dN_q^2)$ & \begin{tabular}{l}$d$: hidden size\\$G$: number of groups\\$K$: voxels in each group\\$N_q$: number of query tokens\end{tabular} & \begin{tabular}{l}$d = 1000$\\$G = 500$\\$K = 30$\\$N_q = 200$\end{tabular}\\
\midrule
MindLLM &$0.0181$ & $O(dN_qN)$ & \begin{tabular}{l}$d$: hidden size\\$N$: number of voxels\\$N_q$: number of query tokens\end{tabular} & \begin{tabular}{l}$d = 100$\\$N = 15000$\\$N_q = 100$\end{tabular}\\
\midrule
\midrule
LLM &$0.0621$ & - & - & -\\
     \bottomrule
    \end{tabular}
\end{table}

\section{Qualitative Analysis}
\label{app:qual}

\begin{figure}
    \centering
    \includegraphics[width=\linewidth]{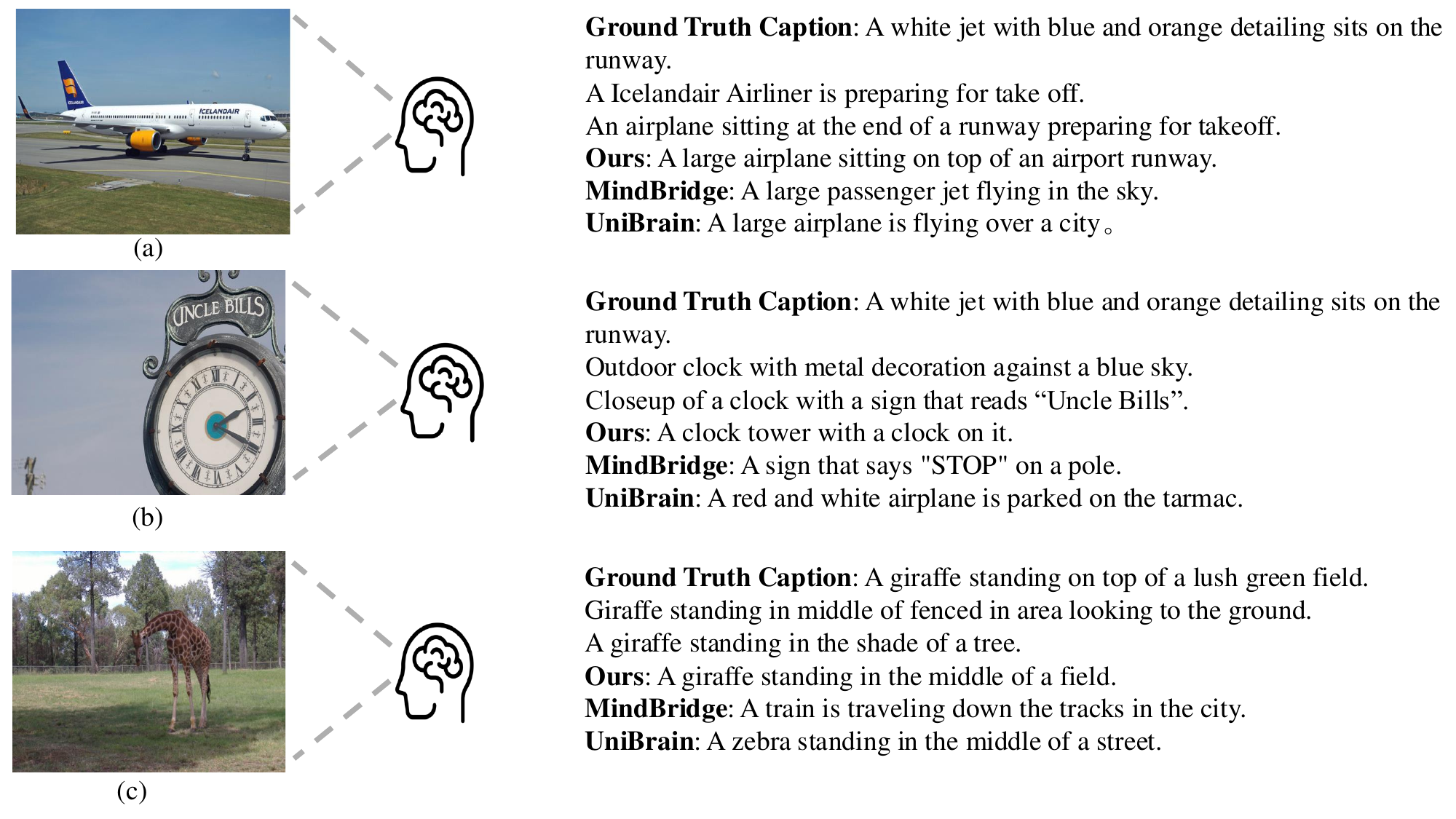}
    \caption{Qualitative Analysis of COCO Captioning. 
    }
    \label{fig:case1}
\end{figure}

\begin{figure}
    \centering
    \includegraphics[width=\linewidth]{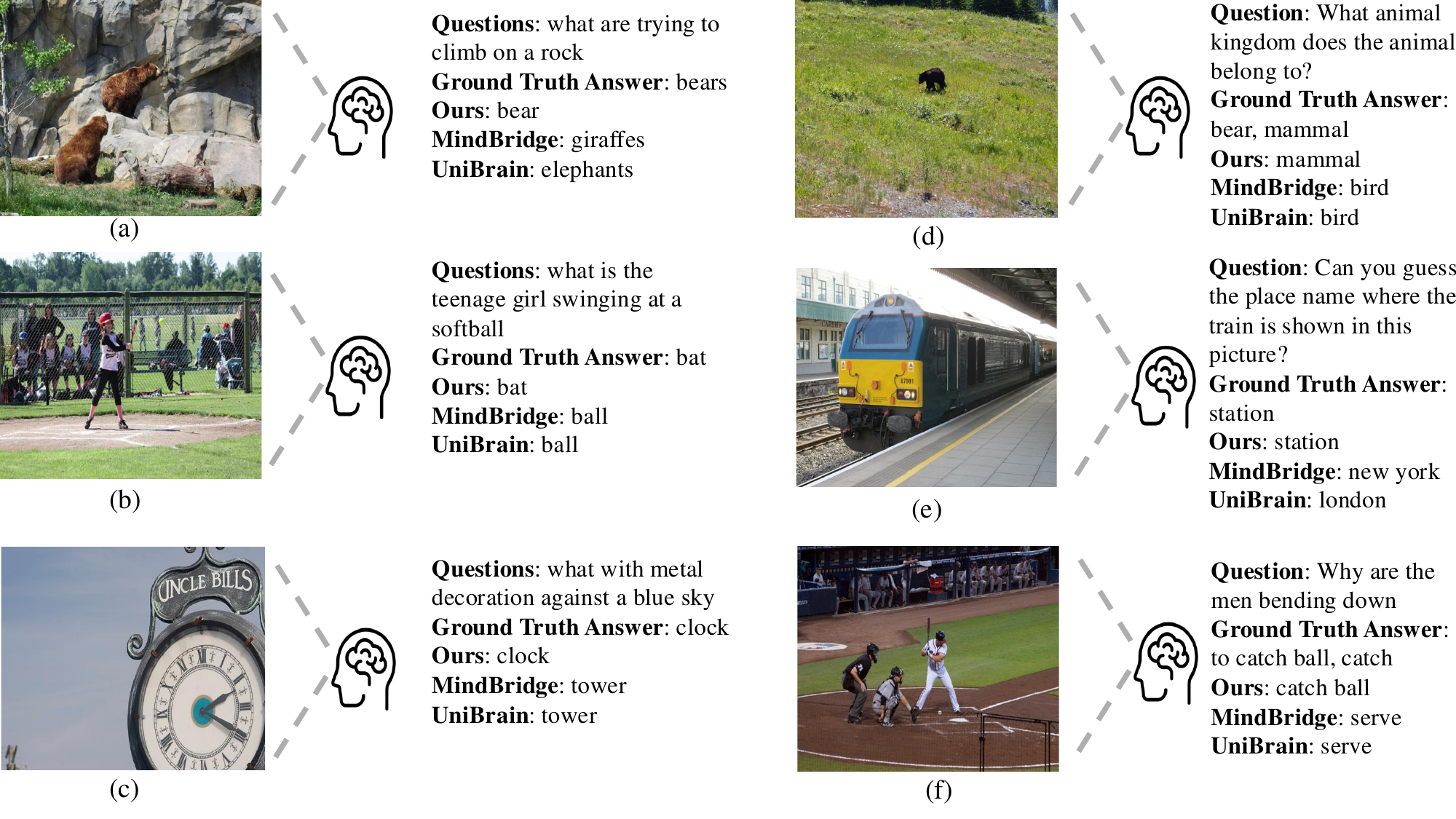}
    \caption{Qualitative Analysis of COCO QA (left column) and OK-VQA (right column).}
    \label{fig:case2_3}
\end{figure}

In this section, we present a qualitative analysis of our model on COCO Captioning, COCO-QA, and OK-VQA, comparing its performance against MindBridge \cite{wang2024mindbridge} and UniBrain \cite{wang2024unibrain}. As shown in Figure \ref{fig:case1} and \ref{fig:case2_3}, our results demonstrate significant improvements in visual understanding across multiple tasks. The model shows strength in the following areas:
\textbf{Static Object Recognition. }The model demonstrates superior accuracy in identifying stationary objects. In comparison with baseline models (MindBridge and UniBrain), our approach shows improvement in spatial context understanding. For example, when analyzing aircraft imagery (e.g., (a) of Figure \ref{fig:case1}), our model correctly identifies "airplane sitting on the runway" while baselines incorrectly interpret the scene as "flying in the sky" or "flying over a city", demonstrating better state-space recognition. \textbf{Action Recognition. }Our proposed model exhibits enhanced capability in distinguishing between similar actions. In sports scenarios (e.g., (f) of Figure \ref{fig:case2_3}), our model correctly identifies "catch ball" while both baselines incorrectly predict "serve", indicating improved action-state discrimination.  \textbf{Potential of neuroscience application. }The demonstrated improvements in object understanding and action recognition suggest the potential for advancing brain-computer interface technology and neural processing research. The model's enhanced capabilities in distinguishing object states and actions could lead to more effective neural prosthetics and improved assistive technologies for individuals with visual or motor impairments.

\section{Ablation Studies on Downstream Tasks}
\label{app:ablation}
Table~\ref{tab:ablation_downstream} reports the results of the ablation study on each downstream task. The results validate the design of neuroscience-informed attention.

\begingroup
\sisetup{
  table-format=3.2,  
  table-align-text-pre=false,
  table-number-alignment=center,
detect-weight=true,detect-inline-weight=math
}
\def\Uline#1{#1\llap{\uline{\phantom{#1}}}}

\begin{table}[htbp]
\vspace{-1em}
\centering
\caption{Ablation study of the key embedding design. Pos Enc. stands for positional encoding. Reg Enc. stands for multiple region encodings. \textit{(x, y, z)+MLP} means we employ an MLP to map the coordinates to the embeddings instead of positional encoding. Note that unlike in Table 2, where we fine-tune the model on each task individually, here we do not fine-tune the model on each individual task and report the performances after brain instruction tuning.}
\label{tab:ablation_downstream}
\vskip 0.1in
\resizebox{0.9\linewidth}{!}{

\begin{tabular}{ccSSSSS}
\toprule
                                   &            & {$(x, y, z)$ + MLP} & {Pos Enc. + fMRI} & {Pos Enc.}       & {Reg Enc.} & {Pos Enc. + Reg Enc.} \\
\midrule      
COCO-QA                            & Accuracy $\uparrow$   & 40.57     & 43.14           & \bfseries 46.29          & 43.90     & \Uline{44.76}               \\
\midrule      
VG-QA                              & Accuracy $\uparrow$  & 20.29     & 20.39           & \bfseries 21.12 & 20.83    & \bfseries 21.12      \\
\midrule      

VQA-v2                             & Accuracy  $\uparrow$ & 45.25     & 46.06           & 47.30           & \Uline{47.44}    & \bfseries 47.86     \\
\midrule      

A-OKVQA                            & Accuracy $\uparrow$  & \Uline{46.02}     & 44.25           & 45.13          & 42.48    & \bfseries 46.90       \\
\midrule      

ST-VQA                             & ANLS  $\uparrow$     & \bfseries 13.70      & 9.70             & \Uline{12.79}          & 10.61    & 11.64               \\
\midrule      


\multirow{2}{*}{Tally QA}                        & Accuracy $\uparrow$  & 44.63     & 49.77           & \Uline{52.29}          & 48.84    & \bfseries 53.01      \\
                                   & RMSE $\downarrow$    & 2.15      & 1.83            & \Uline{1.75}           & 1.96     & \bfseries 1.73       \\
\midrule      

\multirow{8}{*}{COCO-Caption}      & BLEU-1 $\uparrow$    & 49.39     & 54.86           & \Uline{57.03}          & 52.53    & \bfseries 58.98      \\
                                   & BLEU-2  $\uparrow$   & 26.30      & 33.84           & \Uline{37.12}          & 33.00       & \bfseries 39.08      \\
                                   & BLEU-3  $\uparrow$    & 14.06     & 20.67           & \Uline{23.63}          & 20.69    & \bfseries 25.56      \\
                                   & BLEU-4  $\uparrow$   & 8.19      & 13.42           & \Uline{15.51}          & 13.93    & \bfseries 17.36      \\
                                   & METEOR $\uparrow$    & 12.31     & 14.97           & \Uline{16.41}          & 14.81    & \bfseries 17.12      \\
                                   & ROUGE $\uparrow$     & 35.45     & 40.17           & \Uline{42.14}          & 39.62    & \bfseries 43.34    \\
                                   & CIDEr $\uparrow$     & 18.26     & 32.15           & \Uline{41.32}          & 34.25    & \bfseries 46.17       \\
                                   & SPICE $\uparrow$     & 3.94      & 7.14            & \Uline{8.56}           & 6.76     & \bfseries 9.24       \\
\midrule      

\multirow{6}{*}{Paragraph Caption} & BLEU-1 $\uparrow$    & 24.13     & 24.57           &  \Uline{25.90}           & 24.74    & \bfseries 27.17                \\
                                   & BLEU-2 $\uparrow$    & 11.88     & \Uline{13.03}           & 12.41          & 12.61    & \bfseries 14.50                \\
                                   & BLEU-3 $\uparrow$    & 6.12      & 7.16             & 6.10            & \Uline{7.20}     & \bfseries 7.99               \\
                                   & BLEU-4 $\uparrow$    & 3.55      & 4.24            & 3.22           & \Uline{4.35}     & \bfseries 4.67                \\
                                   & METEOR $\uparrow$    & 11.19     & 11.78           & 9.05           & \Uline{12.12}    & \bfseries 14.43               \\
                                   & CIDEr  $\uparrow$    & 3.32      & 2.35            & \bfseries 7.46           & \Uline{4.97}     & 4.30                 \\
\midrule      

\multirow{8}{*}{VQA-E}             & Accuracy $\uparrow$       & 46.74     & 48.17           & \Uline{48.60}          & 48.48    & \bfseries  48.95               \\
                                   & BLEU-1 $\uparrow$    & 35.64     & 36.19           &  \Uline{36.46}          & 35.06    & \bfseries 36.65               \\
                                   & BLEU-2 $\uparrow$    & 18.65     & 19.38           &  \Uline{19.53}        & 18.32    & \bfseries 19.58                \\
                                   & BLEU-3 $\uparrow$    & 10.85     & 11.40            & \bfseries 11.62         & 10.54    & \Uline{11.57}               \\
                                   & BLEU-4  $\uparrow$   & 6.79      & 7.12            &  \Uline{7.29}           & 6.38     & \bfseries 7.31                 \\
                                   & CIDEr  $\uparrow$    & 83.18     & 88.14           &\bfseries 89.24        & 85.34    & \Uline{88.96}               \\
                                   & METEOR $\uparrow$    & 14.20      & 14.81           & \bfseries 15.08         & 14.54    & \bfseries 15.08                  \\
                                   & ROUGE $\uparrow$     & 33.81     & \Uline{35.15}           & \bfseries 35.17        & 34.28    & 34.76    \\

\midrule
\midrule
Average & Perplexity $\downarrow$ & 6.39 & 5.82 & \Uline{5.47} & 5.73 & \bfseries 5.38 \\

\bottomrule

\end{tabular}
}

\end{table}
\endgroup

\section{Architecture Comparison}
\label{app:arch_compare}
We make a side-by-side comparison between our model and related baselines in Table~\ref{tab:arch_tick}. Our model is the only one that achieves the four properties listed in the table's head.

\begin{table}[htbp]
    \centering
    \begin{tabular}{c|cccc}
    \toprule
        Method & cross-subject & unified architecture & shared parameters & using all voxels\\
    \midrule
        MindEye \cite{scotti2024reconstructing} & \xmark & \xmark & \xmark & \cmark \\
        MindEye2 \cite{scotti2024mindeye2} & \cmark & \xmark & \xmark & \cmark \\
        UMBRAE \cite{xia2024umbrae}  & \cmark & \xmark & \xmark & \cmark \\
        MindBridge \cite{wang2024mindbridge} & \cmark & \cmark & \xmark & \xmark \\
        UniBrain \cite{wang2024unibrain} &\cmark & \cmark & \cmark & \xmark \\
        MindLLM & \cmark & \cmark & \cmark & \cmark \\
    \bottomrule
    \end{tabular}
    \caption{Comparison of our model and baselines. Concepts of properties: \textit{cross-subject}: the model performs across subjects; \textit{unified architecture}: the architecture for each subject is the same (the parameters might be different, e.g., MLPs with the same size but different parameters); \textit{shared parameters}: The model uses the same parameters for all subjects; \textit{using all voxels}: the model uses all voxels as inputs, thereby mitigating the risk of information loss from voxel omission. Refer to Section 2.2 and Figure 2 for details of this property.}
    \label{tab:arch_tick}
\end{table}

\section{Experiments on Other Subjects}
\label{app:breakdown}
We report experiments on subjects 2, 5, 7 in Table~\ref{tab:breakdown}. We chose these subjects because they are commonly used in previous works. We observe consistent performance gains over the baselines in the majority of scenarios.

\begingroup
\sisetup{
  table-format=3.2,  
  table-align-text-pre=false,
  table-number-alignment=center,
detect-weight=true,detect-inline-weight=math
}
\begin{table}[htbp]\centering
\caption{Breakdown performances on subjects 2, 5, 7 respectively.}
\label{tab:breakdown}
\vskip 0.1in
\resizebox{\linewidth}{!}{
\begin{tabular}{ccSSSSSSSSSSSSSSS}\toprule
& &\multicolumn{4}{c}{subject 2} & &\multicolumn{4}{c}{subject 5} & &\multicolumn{4}{c}{subject 7} \\
\cmidrule{3-6}\cmidrule{8-11}\cmidrule{13-16}
& &{UMBRAE} &{UniBrain} &{MindBridge} &{MindLLM} & &{UMBRAE} &{UniBrain} &{MindBridge} &{MindLLM} & &{UMBRAE} &{UniBrain} &{MindBridge} &{MindLLM} \\
\midrule
COCO-QA &Accuracy $\uparrow$ &26.85 &47.14 &49.04 &\bfseries 51.71 & &29.86 &45.52 &50.57 &\bfseries 53.14 & & 25.78 & 45.05 & 50.19 & \bfseries 50.76 \\\midrule
VG-QA &Accuracy $\uparrow$ &18.65 &23.53 &24.97 &\bfseries25.96 & &19.38 & 23.65& 26.11 &\bfseries 26.89 & & 18.44 & 23.02 & 25.16 &\bfseries 25.92\\\midrule
VQA-v2 &Accuracy $\uparrow$ &49.96 & 48.17 &51.28 &\bfseries53.17 & &51.49 & 48.81 &51.78 &\bfseries 53.09 & & 50.13 & 47.97 & 50.60 & \bfseries 52.70 \\\midrule
A-OKVQA &Accuracy $\uparrow$&40.54 &43.36 &44.25 &\bfseries49.56 & &43.24 &45.13 &50.44 &\bfseries50.74 & & 40.54 & 46.02 & 45.13 & \bfseries46.90 \\\midrule
ST-VQA &ANLS $\uparrow$&6.61 &9.34 &10.49 & \bfseries 10.85 & &4.31 &9.77 & \bfseries 10.49 & 9.63 & & 5.46 & 8.26 & 13.48 &\bfseries 13.71 \\\midrule
OK-VQA &Accuracy $\uparrow$ &20.70 &32.82 &33.50 & \bfseries 35.73
& &22.98 &30.43 &34.36 & \bfseries 35.04  & & 17.89 & 34.19 & 34.54 & \bfseries 34.87\\\midrule
\multirow{2}{*}{TallyQA} &Accuracy $\uparrow$ &42.21 &55.99 &.20 & \bfseries57.84 & &42.69 &54.65 &58.71 &\bfseries59.3  & & 40.95 & 53.98 & 56.20 & \bfseries 58.41\\
&RMSE $\downarrow$ &4.85 &1.75 &\bfseries1.65 &1.66 & &4.86 &1.64 &\bfseries1.60 & \bfseries1.60  & & 3.65 &1.72 & \bfseries1.66&1.73\\\midrule
\multirow{6}{*}{Paragraph Caption} &BLEU-1 $\uparrow$ &29.29 &27.53 &27.19 &\bfseries30.13 & &29.21 &29.30 &28.70 &\bfseries34.54 & & 29.61 & 25.54 & 27.45 & \bfseries 29.54\\
&BLEU-2 $\uparrow$ &13.82 &14.73 &14.75 &\bfseries16.29 & &14.37 &15.38 &15.21 &\bfseries19.11 & & 14.21& 13.26&14.44 &\bfseries 16.19\\
&BLEU-3 $\uparrow$ &6.44 &8.42 &8.70 &\bfseries11.69 & &6.84 &8.73 &8.87 &\bfseries10.11 & & 6.48 &7.42 &8.05 &\bfseries 9.34\\
&BLEU-4 $\uparrow$ &3.08 &5.10 &5.46 &\bfseries5.96 & &3.28 &5.23 &5.53 &\bfseries7.60 & & 2.95& 4.52 & 4.84 &\bfseries 5.73\\
&METEOR $\uparrow$&11.94 &12.31 &12.36 &\bfseries13.06& &12.68 &11.63 &12.63 &\bfseries13.09 & &12.47 & 11.31 &12.57 &\bfseries 13.54\\
&CIDEr $\uparrow$&5.40 &7.53 &.21 &\bfseries11.90 & &7.46 &6.00 &8.63 &\bfseries12.97 & &\bfseries 8.84 & 4.03 & 5.62 & 8.53\\\midrule
\multirow{8}{*}{VQA-E} &Accuracy $\uparrow$ &47.62 &50.03 &52.30 &\bfseries54.92 & &49.06 &50.18 & 54.21 & \bfseries55.75 & & 48.95& 49.48 & 53.84 & \bfseries 54.68\\
&BLEU-1 $\uparrow$ &30.05 &36.63 &37.48 &\bfseries38.13 & &30.55 &37.41 &38.07 &\bfseries38.40 & & 29.58& 36.71&38.33 & \bfseries 38.41\\
&BLEU-2 $\uparrow$ &14.87 &19.60 &20.16 &\bfseries20.77 & &14.99 &19.96 &20.90 &\bfseries21.04 & &14.43 &19.78 &20.90 &\bfseries 20.98\\
&BLEU-3 $\uparrow$ &8.31 &11.49 &11.76 &\bfseries12.34 & &8.30 &11.66 &12.54 &\bfseries12.60 & & 7.97 & 11.80 & 12.49& \bfseries 12.56\\
&BLEU-4 $\uparrow$&4.99 &7.18 &7.26 &\bfseries7.74 & &4.93 &7.15 &7.94 &\bfseries8.17 & & 4.68 & 7.53 & \bfseries{7.99} & 7.90\\
&CIDEr $\uparrow$ &64.60 &89.50 &91.93 &\bfseries97.68 & &66.47 &89.91 &95.30 &\bfseries99.48 & &61.34 & 90.49&98.23 & \bfseries 99.36\\
&METEOR $\uparrow$ &12.37 &15.12 &15.56 &\bfseries16.02 & &12.51 &15.37 &15.88 &\bfseries16.28 & & 12.09 & 15.02& 16.07 & \bfseries16.16\\
&ROUGE $\uparrow$ &28.49 &35.10 &35.86 &\bfseries36.52 & &28.75 &35.42 &36.43 &\bfseries36.82 & & 28.04 & 35.24& 36.61 & \bfseries36.73\\\midrule
\multirow{8}{*}{FSVQA} &VQA Acc. $\uparrow$ &16.37 &45.62 &48.62 &\bfseries50.19 & &16.40 &45.66 &49.45 &\bfseries50.76 & & 16.37 & 46.11 & 49.19 & \bfseries 51.13\\
&FSVQA Acc. $\uparrow$ &0.00 &40.13 &43.57 &\bfseries45.21 & &0.00 &45.66 &44.60 &\bfseries45.53 & & 0.00 & 40.07 & 44.70 & \bfseries46.07 \\
&BLEU-1 $\uparrow$ &18.98 &86.03 &87.28 &\bfseries87.84 & &19.00 &85.99 &87.76 &\bfseries87.89 & & 18.98& 85.48& 87.29& \bfseries88.05\\
&BLEU-2 $\uparrow$ &9.11 &81.66 &83.14 &\bfseries83.94 & &11.94 &81.52 &83.25 &\bfseries83.87 & & 10.89 & 81.00 & 83.42 & \bfseries84.16\\
&BLEU-3 $\uparrow$ &1.98 &77.61 &79.31 &\bfseries80.28 & &3.42 &77.34 &79.64 &\bfseries80.21 & & 2.82 & 76.84 & 79.70 & \bfseries80.60\\
&BLEU-4 $\uparrow$ &0.00 &73.45 &75.42 &\bfseries76.51 & &1.65 &73.05 &75.99 &\bfseries76.45 & & 0.00& 72.56& 76.11 &\bfseries77.01\\
&METEOR $\uparrow$&5.65 &48.06 &49.26 &\bfseries49.88 & &5.67 &48.01 &49.57 &\bfseries50.10 & & 5.66 &47.86 &49.74 & \bfseries50.18\\
&CIDEr $\uparrow$&3.19 &647.62 &671.26 &\bfseries681.94 & &3.26 &648.21 &675.99 &\bfseries684.12& &3.37 &643.44 & 676.50 & \bfseries684.72\\
\bottomrule
\end{tabular}
}

\end{table}

\endgroup

\section{Latent Space Visualization}
\label{app:latent}
\begin{figure}[htbp]
    \centering
    \includegraphics[width=\linewidth]{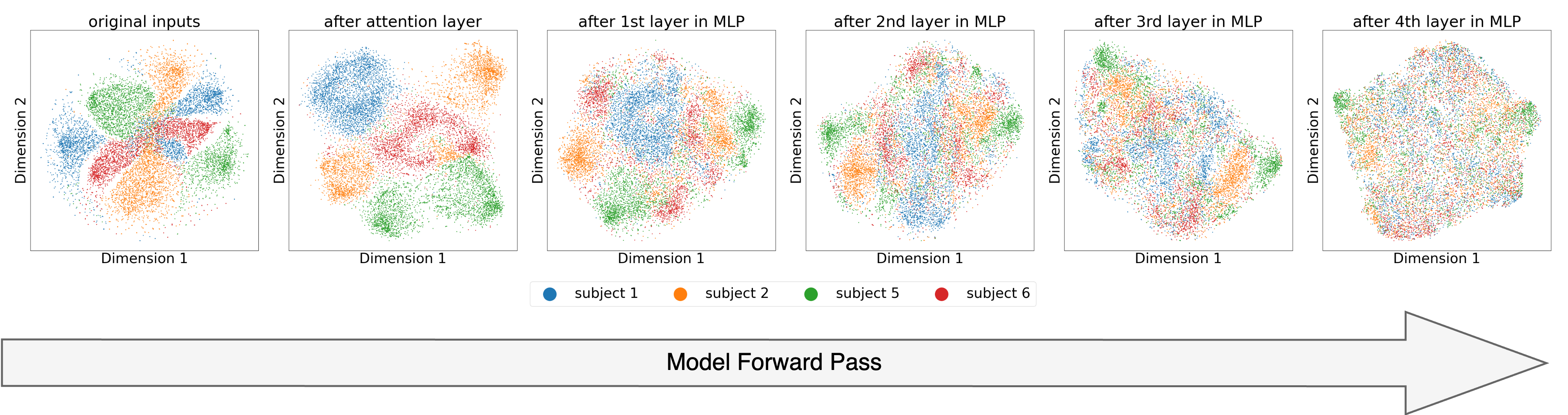}
    \caption{Visualization of latent embeddings across model layers using T-SNE.}
    \label{fig:latent}
\end{figure}

To assess how subject-specific and subject-agnostic information evolves through the model, we visualize the latent embeddings at various stages of the encoder in Figure~\ref{fig:latent}. The original inputs exhibit clear subject-specific clusters. After passing through the neuroscience-informed attention layer, the embeddings remain separable by subject, indicating that subject identity is still preserved. As the data flows through successive layers of the MLP, the subject-specific patterns become increasingly mixed, ultimately forming a subject-agnostic representation. This transformation facilitates generalization on held-out subjects in a zero-shot manner.

\section{Brain Attention Maps of Other Subjects}
\begin{figure}[htbp]
    \centering
    \includegraphics[width=0.8\linewidth]{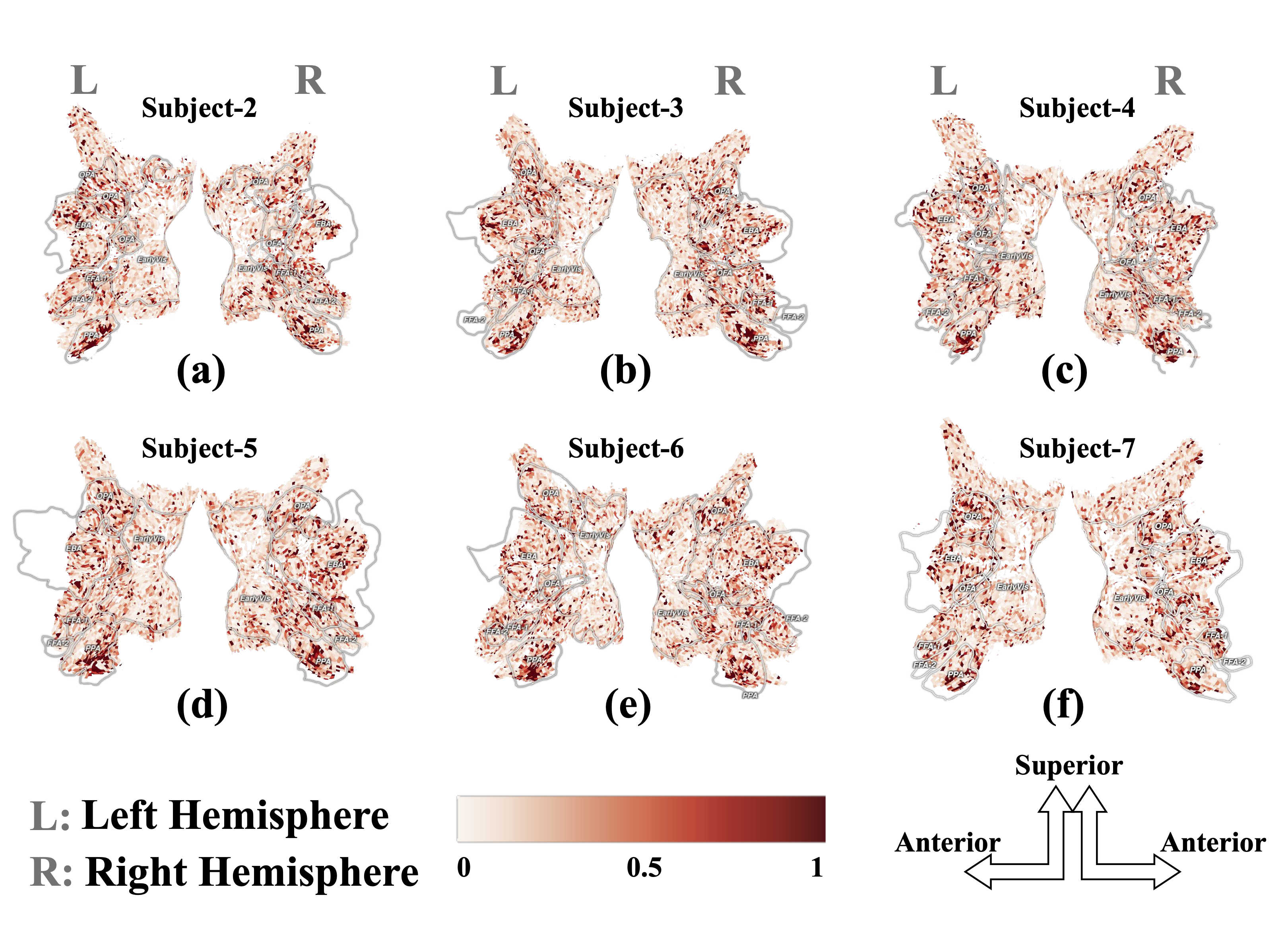}
    \caption{Figure \ref{fig:query}a's query token on different subjects. It shows similar patterns across subjects, suggesting shared brain organization, while also capturing some subject-specific variances.}
    \label{fig:subj2to7}
\end{figure}

We visualize the flat brain attention maps (corresponding to Figure \ref{fig:query}a) in Figure \ref{fig:subj2to7}. We observe that the attention maps exhibit highly similar spatial patterns across subjects, indicating that the model captures shared structural-functional correspondences across human brains, which is expected due to the overall anatomical and functional similarity among individuals. At the same time, we do observe moderate subject-specific variations in the attention maps. These reflect differences in voxel-level functional signals and individual variability in precise voxel locations. The model is able to accommodate these differences through flexible attention mechanisms.

\label{app:brainmap_other}

\end{document}